\documentclass[11pt]{article}

\usepackage[final]{acl}
\usepackage{comment}
\usepackage{times}
\usepackage{latexsym}
\usepackage{amsmath}
\usepackage[T1]{fontenc}
\usepackage[utf8]{inputenc}

\usepackage{microtype}

\usepackage{inconsolata}

\usepackage{graphicx}

\usepackage{hyperref}
\usepackage{url}
\usepackage{enumitem}
\usepackage{booktabs}
\usepackage{multirow}

\usepackage{pifont}
\usepackage{graphicx}

\newcommand{\toolname}{\textsc{PR$^2$}}

\newif\ifshowcomments
\showcommentsfalse
\ifshowcomments
\newcommand{\wcp}[1]{\textcolor{cyan}{[chengpeng: #1]}}
\newcommand{\wac}[1]{\textcolor{cyan}{[wuqi: #1]}}

\newcommand{\xw}[1]{\textcolor{cyan}{[xuwei: #1]}}
\newcommand{\shi}[1]{\textcolor{blue}{[shi: #1]}}
\else
\newcommand{\wcp}[1]{}
\newcommand{\wac}[1]{}
\newcommand{\xw}[1]{}
\newcommand{\shi}[1]{}
\newcommand{\xyz}[1]{}
\fi

\title{Raw Pointer Rewriting with LLMs for Translating C to Safer Rust}

\author{
\textbf{Yifei Gao}$^{1}$, \textbf{Chengpeng Wang}$^{1}$, \textbf{Pengxiang Huang}$^{2}$, \\
\textbf{Xuwei Liu}$^{1}$, \textbf{Mingwei Zheng}$^{1}$, \textbf{Xiangyu Zhang}$^{1}$ \\
$^{1}$Purdue University \quad $^{2}$University of Rochester \\
{\texttt{\{gao749, wang6590, liu2598, zheng618, xyzhang\}@purdue.edu}} \\
{\texttt{pengxiang.huang@rochester.edu}}
}

\usepackage{tcolorbox}
\usepackage{xcolor}

\tcbuselibrary{skins, breakable}

\newtcolorbox{promptbox}[1][Prompt]{
  breakable,
  colback=gray!10,
  colframe=gray!50,
  fonttitle=\bfseries,
  title=#1,
  rounded corners,
  boxrule=0.5pt,
  left=6pt, right=6pt, top=4pt, bottom=4pt
}

\begin{document}
\maketitle
\begin{abstract}
There has been a growing interest in translating C code to Rust due to Rust's robust memory and thread safety guarantees. Tools such as \textsc{C2Rust} enable syntax-guided transpilation from C to semantically equivalent Rust code. However, the resulting Rust programs often rely heavily on unsafe constructs, particularly raw pointers, which undermines Rust’s safety guarantees.
This paper aims to improve the memory safety of Rust programs generated by \textsc{C2Rust} by eliminating raw pointers. Specifically, we propose a raw pointer rewriting technique that lifts raw pointers in individual functions to appropriate Rust data structures. Technically, \toolname\ employs decision-tree-based prompting to guide the pointer lifting process. It also leverages code change analysis to guide the repair of errors introduced during rewriting, effectively addressing errors encountered during compilation and test case execution.
We implement \toolname\ and evaluate it using \texttt{gpt-4o-mini} on 28 real-world C projects. It is shown that \toolname\ successfully eliminates 18.57\% of local raw pointers across these projects, significantly enhancing the safety of the translated Rust code. On average, \toolname\ completes the transformation of a project in 5.02 hours, at a cost of \$1.13. Our code is available at \url{https://github.com/bhcsayx/PR2}.
\end{abstract}

\section{Introduction}

% C has long been recognized as a dominant programming language for both system-level and application-level software development, powering critical infrastructure such as operating systems, embedded devices, and security libraries. 
% Despite its ubiquity and efficiency, its lack of built-in memory~\cite{van2021toward,van2023memory,lord2023urgent} and type safety~\cite{yang2018source,turner2014security,avots2005improving} has led to numerous high-impact vulnerabilities. A well-known example is the Heartbleed bug~\cite{heartbleed}, an out-of-bounds read in the OpenSSL library caused by improper memory handling. This flaw resulted in the large-scale leakage of sensitive user data, including passwords and private keys, and caused severe reputational damage to organizations that relied on OpenSSL for secure communication. Such vulnerabilities highlight the inherent risks of using low-level languages such as C that do not enforce strict compile-time checks on resource management.

C has long been a dominant language for system and application software, powering critical infrastructure like operating systems, embedded devices, and security libraries. However, its lack of built-in memory~\cite{van2021toward,van2023memory,lord2023urgent} and type safety~\cite{yang2018source,turner2014security,avots2005improving} has led to significant vulnerabilities. A notable example is the Heartbleed bug, an out-of-bounds read in OpenSSL caused by improper memory handling, which led to large-scale privacy leaks, highlighting the risks of C language lacking strict compile-time checks for resource management.

% \smallskip
% \emph{\textbf{Motivation and Goal.}}
% Motivated by the desire to eliminate dangerous memory handling pitfalls, many developers have adopted Rust~\cite{rustbook}, a systems programming language designed to enforce rigorous memory and type safety guarantees. Rust’s ownership model imposes strict constraints on resource allocation and deallocation~\cite{matsakis2014rust}, while its borrow checker ensures that data races, dangling pointers, and use-after-free errors are statically prevented at compile time. These safety mechanisms enable Rust to achieve performance comparable to C while significantly reducing memory-related vulnerabilities inherent in C programs. As a result, numerous projects (e.g., s2n-tls~\cite{s2ntls}) have undertaken efforts to migrate legacy C codebases to Rust, thereby extending the longevity of critical software systems and strengthening the security of the infrastructure that relies on them.
Driven by the need to eliminate dangerous memory handling issues, many developers have turned to Rust, a systems programming language that enforces strict memory and type safety guarantees. Rust’s ownership model imposes tight constraints on resource allocation and deallocation~\cite{matsakis2014rust}, while its borrow checker prevents various memory errors at compile time. These safety features allow Rust to deliver performance on par with C, while significantly reducing memory-related vulnerabilities. Consequently, many projects (e.g., \texttt{s2n-tls}) are migrating legacy C codebases to Rust, enhancing the security and longevity of critical software.

Unfortunately, translating C code to Rust is challenging. Tools like C2Rust~\cite{c2rust} can achieve syntax-level transpilation but often rely on raw pointers and unsafe functions, which bypasses Rust's safety mechanisms, still introducing memory risks in the Rust code. To generate safer Rust code, existing studies~\cite{emre2021translating, emre2023aliasing, zhang2023ownership, hong2024don} have explored transforming raw pointers into references or other data structures. For example, Laertes~\cite{emre2021translating} and its enhanced version~\cite{emre2023aliasing} use pointer analysis to reason about value lifetimes, enabling conversion to references, while a recent study eliminates output parameters by introducing additional return values~\cite{hong2024don}. However, these static analysis techniques are rule-based, limiting their ability to handle all raw pointers, as will be shown in the motivating example.

\begin{figure*}[t]
	\centering
	\includegraphics[width=\textwidth]{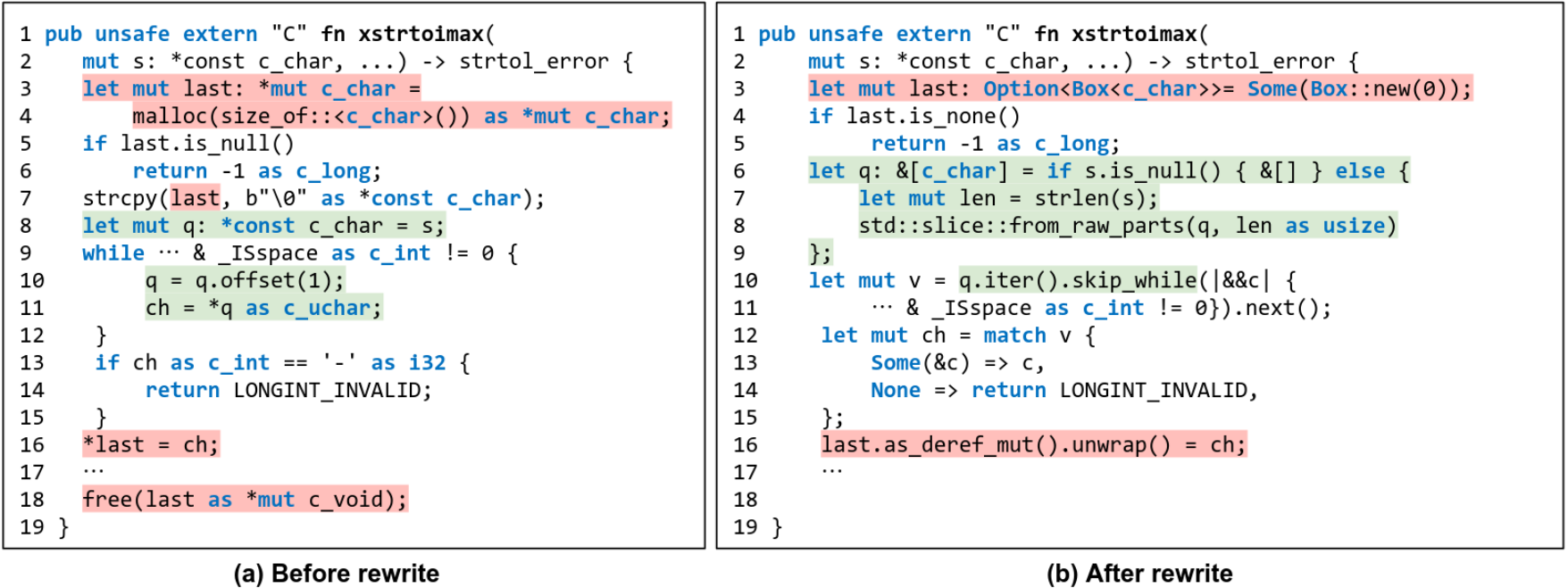}
	\caption{A motivating example of eliminating raw pointers in the tool \textsc{C2Rust}  generated Rust program by rewriting them with Rust data structures, namely a \texttt{Option} of a \texttt{Box} and a slice. (a) The program generated by \textsc{C2Rust} from a C program; (b) A safer Rust program after raw pointer rewriting.
    % \cp{@yifei. (1) Change *last = q[len - 1] to *last = ch. (2) What's the meaning of the while condition at line 10 in Figure~\ref{fig:example}(a)? What does \_ISspace mean?}
    %\cp{@mingwei. The font size of sub-captions should be smaller. In (b), the commented rewrite results should be labeled and shown in other colors.}
    }
    % of transforming raw pointers into safe Rust structures. (a) The original unsafe Rust function uses raw pointers for pointer arithmetics. (b) The rewritten safe Rust function replaces raw pointers with Rust’s ownership-based abstractions, including \texttt{Box}, \texttt{\&[T]}(slices), and \texttt{Option}, ensuring safer memory management and iteration.}
	\label{fig:example}
	\vspace{-5mm}
\end{figure*}

Recent advances in large language models (LLMs) have impacted programming, showing strong abilities in code comprehension~\cite{cui2024code,nam2024using, zheng2025large} and transformation~\cite{yang2024exploring, yuan2024transagent,yin2024rectifier}. Particularly, these models can help translate C code into safer Rust by understanding high-level memory usage intentions hard to obtain through static analysis, guiding the conversion of raw pointers to safer structures like slices. For example, an LLM can recognize one raw pointer used to traverse a memory region and convert it to a slice in Figure~\ref{fig:example}(a). Additionally, LLMs have shown promise in program repair, fixing bugs from compilation errors or test failures, providing a degree of correctness assurance for generated Rust code~\cite{10.1145/3650212.3680323,bouzenia2024repairagent,kulsum2024case}.

Based on the above insights, we propose \toolname, a novel C-to-Rust transpilation technique that combines \textsc{C2Rust} for syntax-level translation with LLMs to eliminate raw pointers, eventually producing safer Rust code. With the guidance of a decision tree, \toolname{} prompts LLMs for the \emph{pointer lifting}, which replaces the low-level raw pointers with safer Rust constructs, such as \texttt{Option}, \texttt{Vec}, \texttt{Box}, slices, and references. Besides, \toolname{} iteratively repairs the statements that use the newly introduced Rust data structures if the rewrite is wrong, ensuring the generated Rust code can pass the memory safety check during the compilation and passes the test cases at runtime. Particularly, we perform the \emph{code change analysis} in the repair stage, which helps direct the LLMs to focus on the statements most likely responsible for errors. While \toolname{} doesn't eliminate all raw pointers, it significantly reduces reliance on them and thus improves memory safety.

In summary, the contributions of this work include: 
(1) We introduce the raw pointer rewriting technique systematically eliminating raw pointers in Rust programs translated from C, facilitating translation to safer Rust.
(2) We develop decision tree-based prompting to lift raw pointers into appropriate Rust data structures, and employ code change analysis to guide program repair in response to compilation errors and test case failures.
(3) We implement \toolname{} and evaluate it on 28 real-world programs ranging from 437 to 136k LOC. Utilizing \texttt{gpt-4o-mini}, it eliminates 2,255 local raw pointers, 3.46 times more than existing techniques, while preserving program semantics in all manually sampled cases, with an average time cost of 5.02 hours and an average financial cost of \$1.13 per project.

\begin{figure*}[t]
	\centering
    \includegraphics[width=\textwidth]{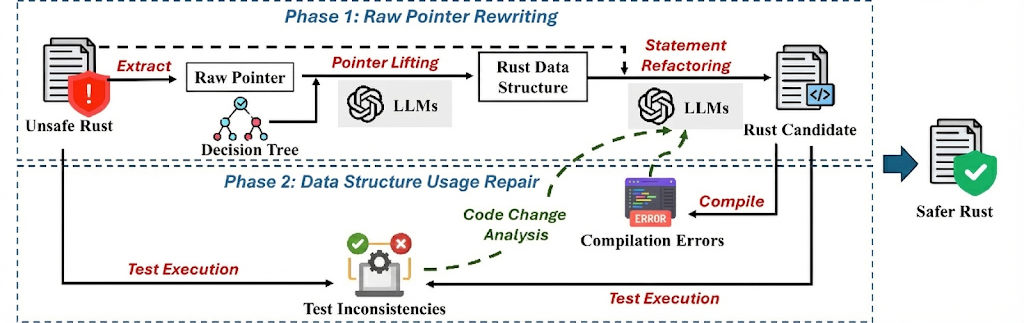}
    %\wcp{TODO: Please change: (1) Peephole Rewriting -> Raw Pointer Rewriting; (2) Peephole repair -> Data Structure Usage Repair}
	\vspace{-5mm}
    \caption{The workflow of \toolname}
    \vspace{-4mm}
    \label{fig:workflow}
\end{figure*}

\section{Related Work}
\label{sec:related}

Rust’s compile-time memory safety has motivated extensive work on automating C-to-Rust translation, broadly divided into two categories. Specifically, the first line of research directly translates C into unsafe but functionally equivalent Rust. \textsc{Corrode}~\cite{corrode} was an early attempt focusing on module-level semantic preservation. \textsc{C2Rust}~\cite{c2rust} remains the most advanced industrial-grade transpiler, offering automated refactoring and cross-checking. \textsc{Citrus}~\cite{citrus} provides syntax-based translation for manual refactoring despite limited semantic preservation. Recent LLM-based approaches~\cite{yang2024vert, shetty2024syzygy} generate multiple candidate translations and use test cases to select correct outputs.
The second category begins with a semantics-preserving Rust baseline and incrementally enhances memory safety by rewriting raw pointers. \textsc{Laertes}~\cite{emre2021translating} optimistically rewrites pointers into references, while its enhanced version~\cite{emre2023aliasing} applies more precise pointer analysis, albeit risking dynamic semantic changes. \textsc{Nopcrat}~\cite{hong2024don} and \textsc{Tymcrat}~\cite{hong2025type} focus on refining function parameters into safe constructs like \texttt{Option} and \texttt{Result}.
Although the semantic equivalence can be guaranteed, existing techniques only support rewriting a restrictive form of raw pointers, leaving numerous raw pointers not rewritten in the Rust code.

With the rapid development of generative AI,
LLMs have greatly advanced cross-language code translation by enabling functionality-preserving conversions. However, LLM-generated translations often suffer from syntax errors and semantic inconsistencies, limiting their practical reliability. To mitigate this, many approaches incorporate test cases to detect errors and iteratively refine outputs.
For example,
\textsc{UniTrans}~\cite{yang2024exploring} generates and executes test cases to validate translations across languages, using feedback for refinement. \textsc{CoTran}~\cite{jana2024cotran} goes further by fine-tuning an LLM with reinforcement learning for Java–Python translation, using compiler feedback and symbolic execution to ensure semantic equivalence.
Despite these advances, C-to-Rust translation poses unique challenges. Rust’s strict ownership, borrowing, and lifetime rules require fundamental changes in memory management and pointer handling, unlike C++, Java, or Python. A recent work~\cite{hong2025type} tackles signature-level type migration by converting C function signatures into idiomatic Rust, but it is limited to interfaces and still reports compiler and semantic errors, highlighting the insufficiency of LLMs alone for Rust’s strict type system.

\section{Approach}
\label{sec:approach}
This section presents technical details of \toolname,
of which the workflow is shown in Figure~\ref{fig:workflow}.
Specifically, \toolname{} attempts to eliminate raw pointers in single functions by using Rust data structures (Phase~1) and repairs the usage of the introduced Rust data structures (Phase 2) to ensure that the generated Rust code can be compiled successfully and pass the test cases.
If the data structure usage repair fails a specific number of times, denoted by $N$, \toolname\ gives up replacing the raw pointer.
In our implementation, we set $N$ to five by default.
The details of raw pointer rewriting and data structure usage repair are demonstrated in Section~\ref{subsec:rewrite} and Section~\ref{subsec:repair}, respectively.

\subsection{Raw Pointer Rewriting}
\label{subsec:rewrite}

This section presents the details of rewriting raw pointers using Rust data structures.
Specifically, we introduce two important stages, i.e., \emph{pointer lifting} and \emph{statement refactoring}, in Section~\ref{subsubsec:pia} and Section~\ref{subsubsec:pr}, respectively.

\subsubsection{Pointer Lifting}
\label{subsubsec:pia}
% \cp{I replace the pointer intention analysis with pointer lifting. The output of this stage is a Rust data structure.}
Existing techniques~\cite{emre2021translating, emre2023aliasing, zhang2023ownership, hong2024don} primarily focus on replacing raw pointers with references based on the results of pointer analysis.
However, due to challenges in analyzing the semantics of complex program constructs, such as loops and library function calls, these approaches often fail to identify opportunities to refactor raw pointers into more expressive Rust data structures, such as \texttt{Vec} and slices.
This limitation degrades the overall effectiveness of these methods, resulting in many raw pointers being retained in the translated Rust programs.

\begin{figure}[t]
	\centering
	\includegraphics[width=\linewidth]{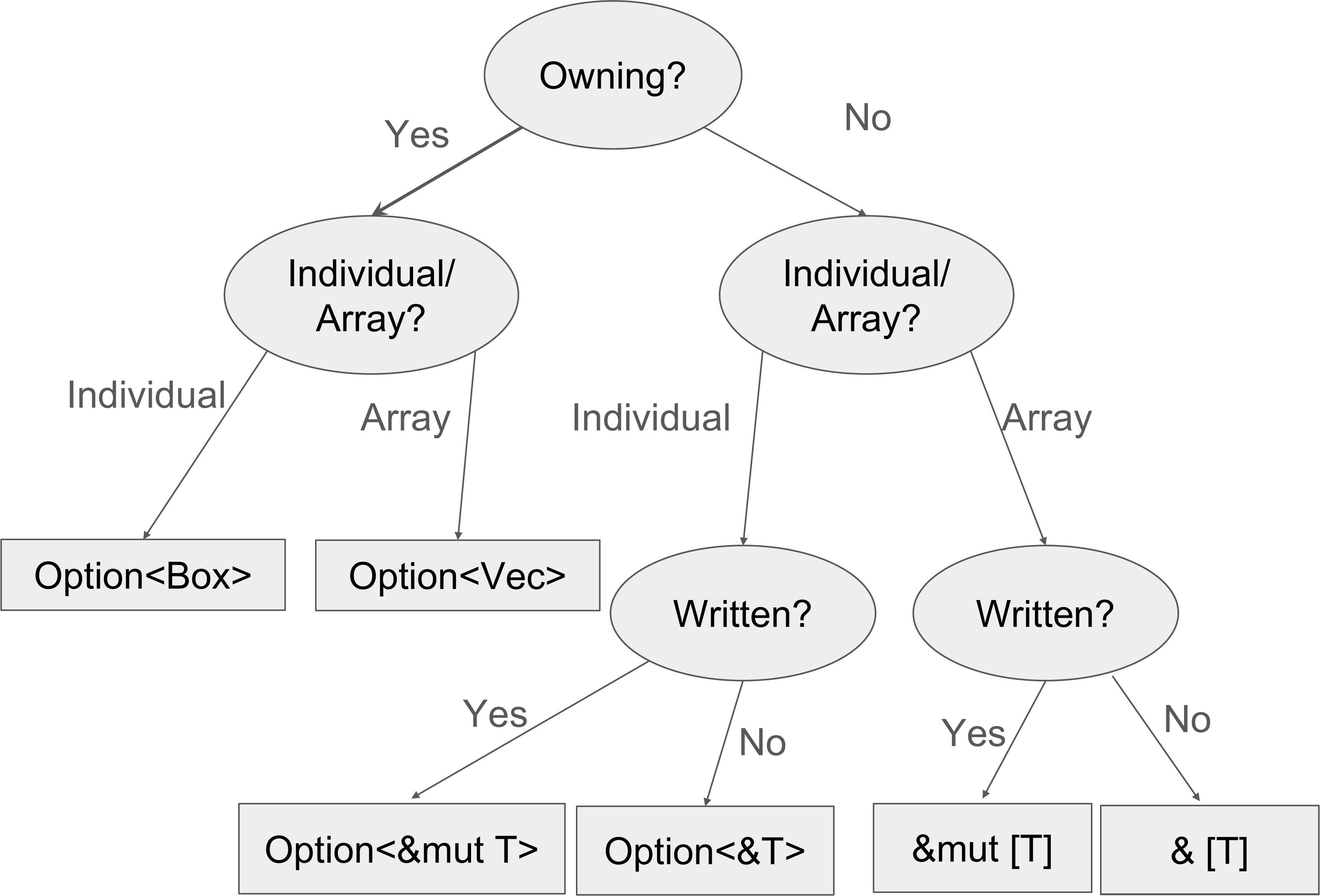}
    % \vspace{-3mm}
	\caption{The decision tree used for the pointer lifting}
	\label{fig:tree}
	% \vspace{-4mm}
\end{figure}

% \vspace{-12mm}

\begin{figure}[t]
	\centering
	\includegraphics[width=\linewidth]{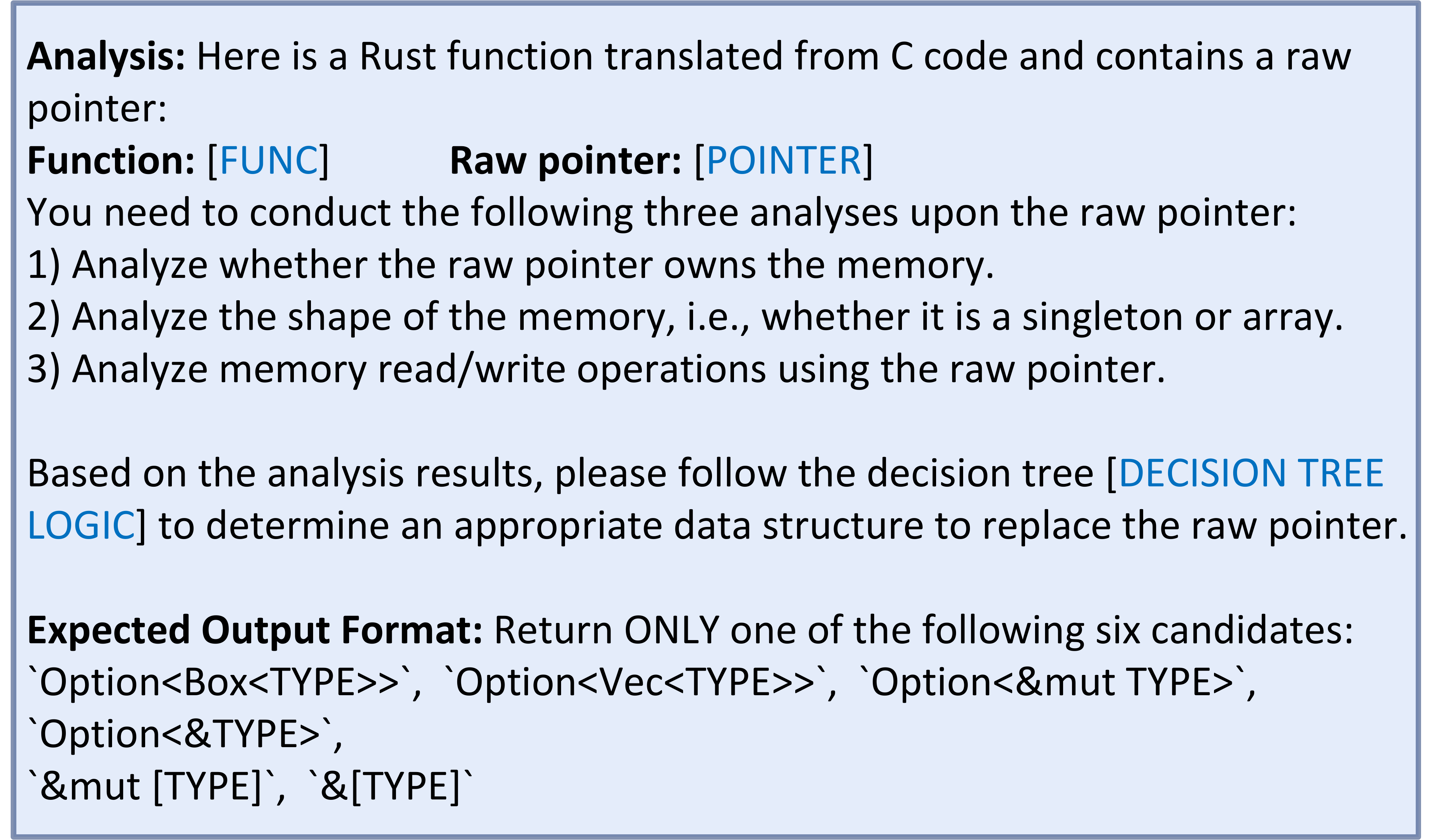}
    % \vspace{-7mm}
	\caption{The prompt template used in the pointer lifting}
	\label{fig:analysis}
	\vspace{-5mm}
\end{figure}

To enable the rewriting of raw pointers using a wider range of Rust data structures, we leverage the LLMs' capability to understand program semantics, that is, \emph{the LLMs can act as a flexible program analyzer when guided by a well-crafted prompt}.
To this end, we introduce a decision tree-based prompting strategy that directs the LLMs to perform semantic analysis on raw pointers.
This analysis helps the LLMs identify suitable safe Rust data structures for replacement.
As illustrated in Figure~\ref{fig:tree}, we enforce the LLMs to reason about three key properties of each raw pointer: \emph{ownership}, \emph{the shape of the memory buffer} (i.e., singleton or array), and \emph{memory access} (i.e., read/write behavior).
These properties collectively determine whether and how a raw pointer can be replaced with a specific data structure.
Concretely, the \underline{ownership property} helps determine whether the pointer can be refactored into a reference, \texttt{Box}, or \texttt{Vec};
the \underline{shape property} indicates whether the pointer corresponds to a single value or a collection (e.g., \texttt{Vec} or slice);
and the \underline{memory-access property} informs whether the resulting data structure should be mutable or immutable.
Since raw pointers can be null, we conservatively wrap all inferred data structures in \texttt{Option}.
Following this decision tree, we can instantiate the prompt template shown in Figure~\ref{fig:analysis} to generate a concrete prompt that guides the LLMs in lifting the raw pointer to an appropriate Rust data structure. As an example, the LLMs can correctly infer the non-owning, array-like, and immutable properties of pointer \texttt{q} (Line 8 in Figure~\ref{fig:example}(a)) and determine the data structure as an immutable slice.

% \cp{Only these six kinds of data structures? Why so many Option cases? I think we should arrange four internal nodes (nullability, owning, shape, written or not). Make the tree as balanced as possible.}

% \begin{example}
% Consider the raw pointer \texttt{q} in Figure~\ref{fig:example}(a). 
% First, the LLMs can infer that \texttt{q} does not own the memory, as the string represented by the buffer originates from the caller of the \texttt{xstrtoimax} function.
% Second, the operations within the loop (lines 9--12)—in particular, the call to the \texttt{offset} function at line 10 and the subsequent dereference at line 11—suggest that \texttt{q} points to an array-like memory region rather than a singleton.
% Third, the LLMs identifies that there are no write operations involving \texttt{q}, implying the memory is read-only.
% According to the decision tree in Figure~\ref{fig:tree}, these properties collectively indicate that \texttt{q} can be refactored into an immutable slice.
% Similarly, we can lift the raw pointer \texttt{last} in Figure~\ref{fig:example}(a) to an \texttt{Option} of \texttt{Box}.
% \end{example}

In contrast to rule-based symbolic static analysis approaches, our decision tree-based prompting strategy enables the LLMs to leverage its semantic understanding and creativity to identify more expressive patterns of raw pointer manipulation, such as those involving loops and library function calls, which are typically difficult for traditional symbolic analysis to handle.
Moreover, our approach is easily extensible: by designing more sophisticated decision trees and corresponding prompts, we can support more complex data structures.
In this work, we focus on sequential data structures for non-singleton values.
As one of the future works, we plan to extend \toolname\ to support additional collection types, such as sets and maps.

% \cp{@yifei. Add three or four sentences after the examples in each technical subsection. Summarize the technical design by highlighting its advantages, e.g., the advantage of peephole design. Section 4.3.1 and Section 4.3.2 are similar.}

% \begin{figure}[t]
% 	\centering
% 	\includegraphics[width=0.5\textwidth]{Fig/reason1.png}
% 	\caption{Reasoning Process of Pointer p}
% 	\label{fig:reason1}
% 	\vspace{-1mm}
% \end{figure}

\subsubsection{Statement Refactoring}
\label{subsubsec:pr}
\begin{figure}[t]
	\centering
	\includegraphics[width=\linewidth]{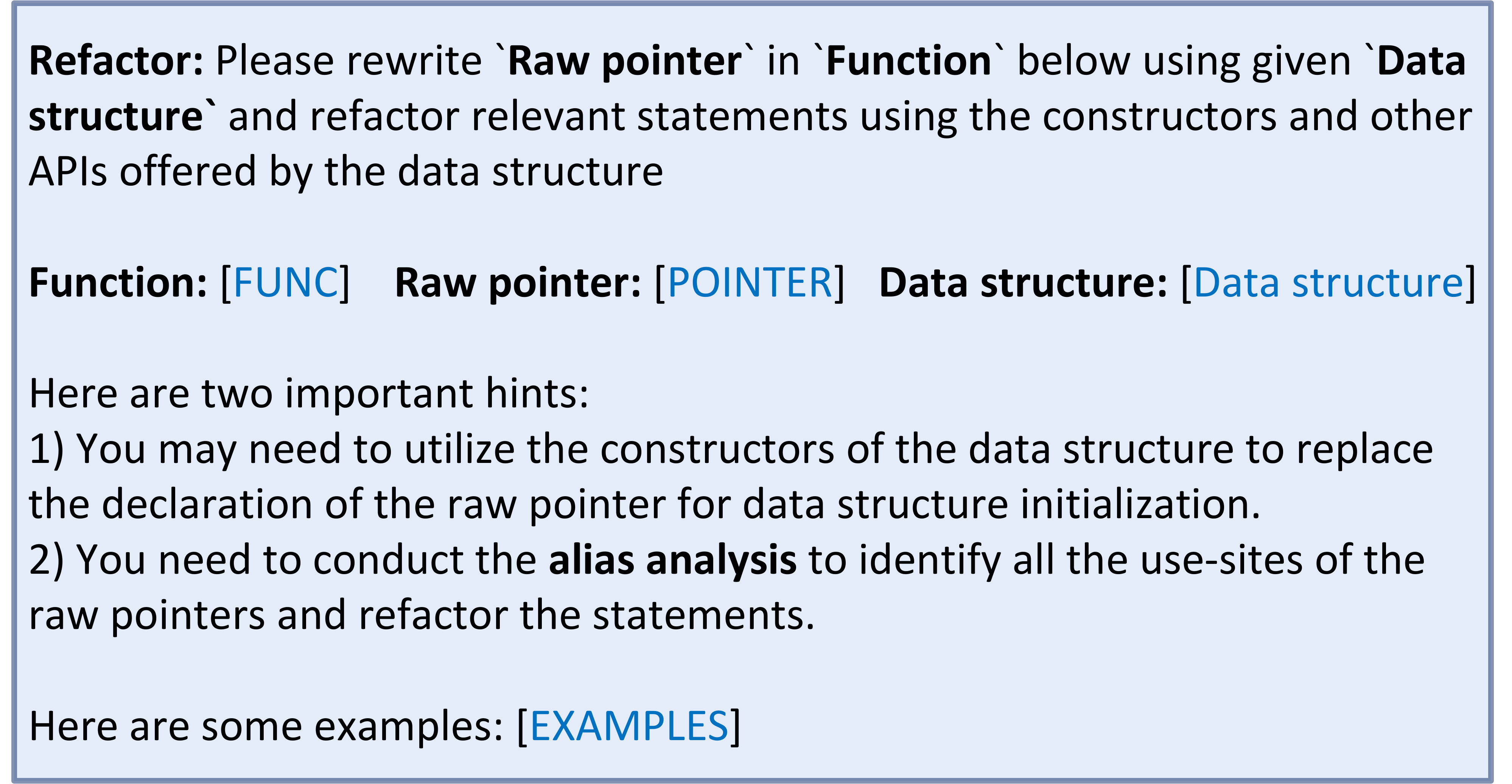}
	\caption{The prompt template for statement refactoring}
	\label{fig:rewrite}
	\vspace{-6mm}
\end{figure}

Based on the result of pointer lifting, we proceed to replace the raw pointer with the suggested data structure.
This process involves refactoring both the declaration of the raw pointer and all statements that make use of it.
As illustrated by the prompt template in Figure~\ref{fig:rewrite}, we provide two key hints to the LLMs.
First, we instruct the LLMs to replace the declaration of the raw pointer with a constructor of the suggested data structure.
Second, we enforce the LLMs to reason about the aliasing facts of the raw pointer in order to locate all potential use-sites,
and refactor them using the appropriate APIs provided by the data structure.
Notably, the statement refactoring leverages two core capabilities of the LLMs, including analyzing semantic properties of the program (e.g., alias analysis) and its ability to generate code (e.g., rewriting statements accordingly), as shown in Figure~\ref{fig:example}(b).

% \begin{example}
% Consider the raw pointer \texttt{q} in the motivating example.
% As the pointer lifting suggests using a slice for refactoring, the LLMs can further identify its use sites within the loop from lines 9 to 12 in Figure~\ref{fig:example}(a), including the offset computation and dereference operations.
% Leveraging its semantic analysis capabilities, the LLMs recognizes that the loop traverses the buffer and skips space characters, which can be elegantly implemented using the APIs \texttt{iter} and \texttt{skip\_while}, as shown in Figure~\ref{fig:example}(b).
% Similarly, the LLMs identifies the declaration of the raw pointer \texttt{last} at lines 3 and 4, its initialization at line 7, its dereference at line 16, and the explicit memory deallocation at line 18 in Figure~\ref{fig:example}(a).
% Using the \texttt{Option} constructor and the API \texttt{as\_deref\_mut}, the LLMs refactors these statements, as shown at lines 3 and 16 in Figure~\ref{fig:example}(b).
% Notably, the LLMs respects Rust’s memory model and omits explicit memory deallocation during the statement refactoring, since wrapping a \texttt{Box} in an \texttt{Option} can facilitate the automatic memory declaration,
% which is enabled by Rule 2 demonstrated in Section~\ref{subsec:ownership}.

% \end{example}

\begin{comment}
Intuitively, our peephole rewritingaims at rewriting the forward slice starting from the definition of raw pointer with the range of the context function. This enables us to perform rewrites for safer Rust within the limit of LLM's context window, also limit the sites of potential compiler and runtime errors to the context function, avoiding errors propagating in the whole program.
% Figure \ref{fig:prexample} demonstrates one example guiding LLMs to rewrite raw pointers into vectors.
\end{comment}

% \begin{figure}[t]
% 	\centering
% 	\includegraphics[width=0.5\textwidth]{Fig/prexample.png}
% 	\caption{Example in Peephole Rewrites Prompt \cp{Please delete this figure. Just use an example for discussion. No need to show the prompt and the responose.}}
% 	\label{fig:prexample}
% 	\vspace{-1mm}
% \end{figure}

\subsection{Data Structure Usage Repair}
\label{subsec:repair}
% \cp{I changed the title of this subsection.}
This section presents data structure usage repair,
which consists of \emph{compilation error fixing} (Section~\ref{subsubsec:comp}) and \emph{testcase error fixing} (Section~\ref{subsubsec:sem}).

\begin{comment}
In this subsection we illustrate how we fix compiler and testcase errors (i.e., . Our core insight for repairing is also repairing in a peephole way, which is based on our observation that even if we focus rewrites of pointers inside single functions, some of them may still be too long (e.g., 150-200 lines) and may include other untouched raw pointers, interfering LLMs for correct repairs. Therefore, for both compiler and testcase errors, we conduct a \emph{\textbf{code change analysis}}, in order to extract information about the location of errors from compiler or testcase feedback, and provide them to LLMs along with function context, guiding them to focus on these locations. For every raw pointer being refactored, we leverage a retries threshold for the fix. We give up rewriting of current pointer if the sum of prompting retries on compilation errors fixing and testcase error fixing exceeds the threshold. We design this threshold since LLMs responses trying to fix testcase errors may introduce new compilation errors and we may need to go back to \texttt{compiler\_check\_fix} to fix them. By default we set this retries threshold to 5. We introduce compilation errors fixing in \ref{subsubsec:comp} and testcase errors fixing in \ref{subsubsec:sem}.
\end{comment}

\subsubsection{Compilation Error Fixing}
\label{subsubsec:comp}
As shown in Section~\ref{subsec:rewrite}, only the code within a single function is modified during the raw pointer rewriting.
This design implies that any resulting compilation error must be introduced by the modified statements within that function,
enabling us to narrow down the scope of fixing to the function for the error fixing.
However, a single function may contain many raw pointers, most of which are unrelated to the repair task.
What is even worse, the function itself can be quite large (e.g., over one hundred lines), with the majority of statements not interacting with the memory addressed by the raw pointers.
These unrelated raw pointers and statements introduce irrelevant tokens as the \emph{noise} into the repair context,
which poses a significant challenge for the LLMs, often leading to hallucinations and repair failures.

\begin{figure}[t]
	\centering
	\includegraphics[width=\linewidth]{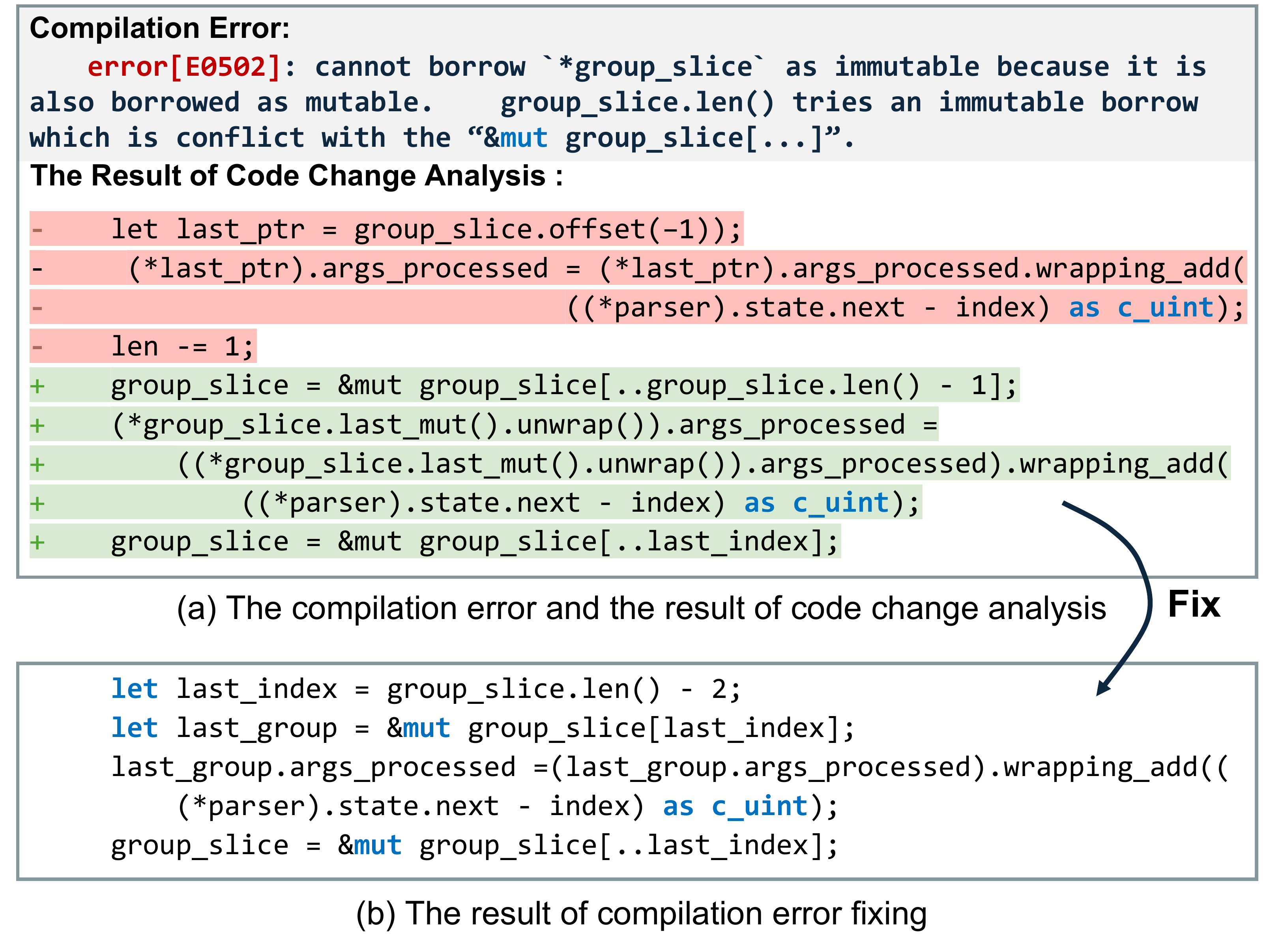}
    \vspace{-8mm}
	\caption{An example of compilation error fixing}
	\label{fig:errfix1}
	\vspace{-6mm}
\end{figure}

To enable more informed repairs using the LLMs,
we filter out irrelevant tokens from the repair context.
Technically, we introduce \emph{code change analysis}, which computes the difference between the function before and after the raw pointer rewriting.
The resulting added, deleted, or modified statements are the most likely sources of compilation errors.
By feeding code changes along with error messages into the LLMs, we can effectively guide the LLMs to focus its \emph{attention} on the most relevant statements, improving its ability to fix compilation errors. Due to space limit, we do not show the prompt template in the paper as the design is very standard based on the code change analysis. Figure~\ref{fig:errfix1} shows an example of using code change analysis to fix a conflicting immutable (\texttt{group\_slice.len()}) and mutable (\texttt{\&mut group\_slice}) borrows error. Specifically, Rust compiler allows only one mutable reference to a variable at any time, thus calling the API \texttt{len} to the slice will raise an error as it needs to hold an immutable reference. To resolve this error, the fix calls the API \texttt{len} before the mutable slice is created, and stores the length value in a new variable, avoiding the interleaving of references.
\subsubsection{Testcase Error Fixing}
\label{subsubsec:sem}

\begin{figure}[t]
	\centering
	\includegraphics[width=\linewidth]{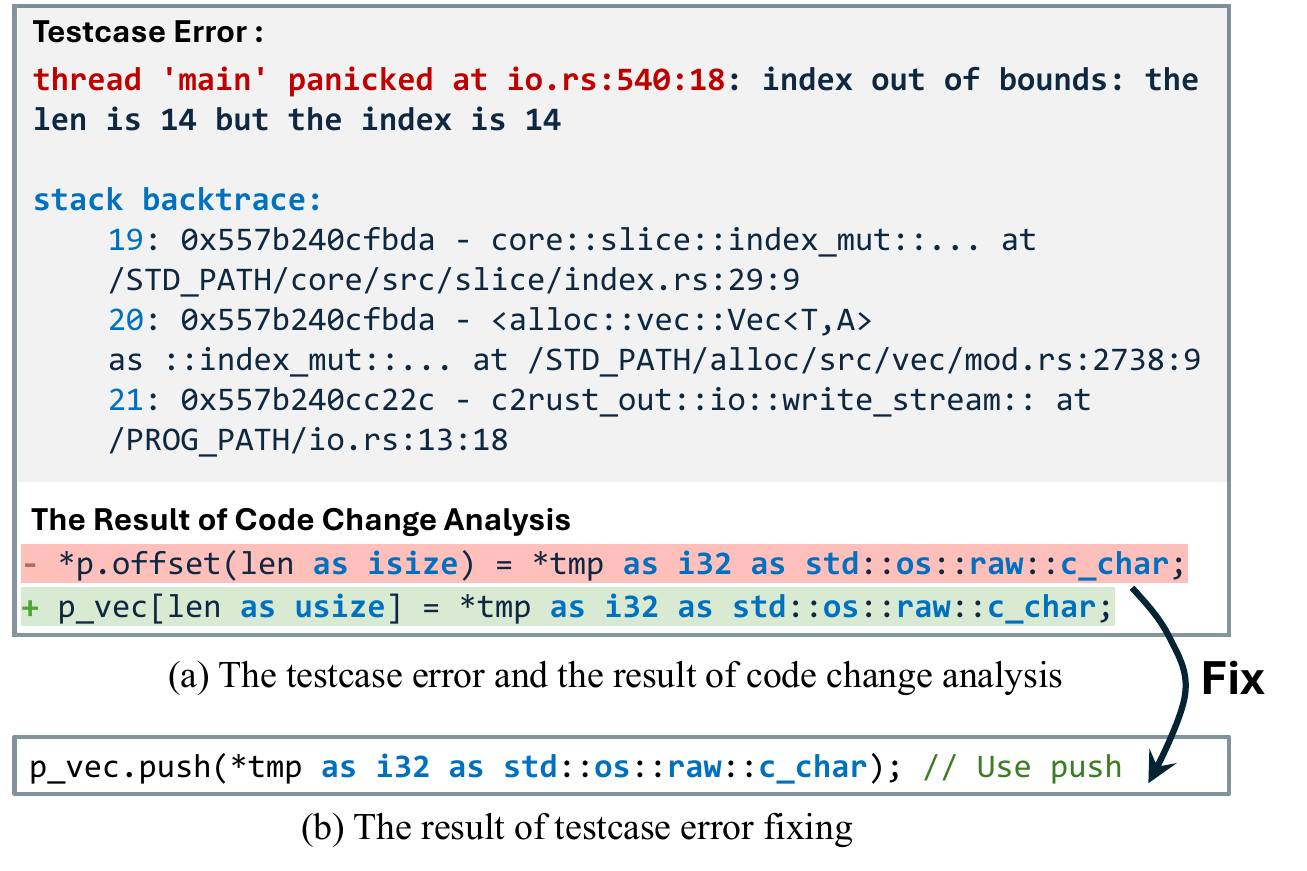}
    \vspace{-8mm}
	\caption{An example of testcase error fixing
    }
	\label{fig:errfix2}
	\vspace{-4mm}
\end{figure}

Once we obtain a Rust program with all compilation errors fixed, we proceed to run the program’s test cases to uncover potential violations of the original input-output relationship.
The overall design of testcase error fixing closely follows that of compilation error fixing described in Section~\ref{subsubsec:comp}. %so we omit further details here.
Figure~\ref{fig:errfix2} shows a prompt template and an example of testcase error fixing.
It is important to note that passing all test cases which the original Rust program pass is a necessary condition for a valid translation, although it does not theoretically guarantee semantic equivalence.
In practice, the validity of our approach depends on the quality of the available testcases.
While many real-world projects include comprehensive test suites, there may be corner cases not captured by the tests.
\begin{table*}[t]\centering
\caption{The main statistics of \toolname\ 
%\wcp{Fix the data in ID 17 and ID 27.}
}\label{tab:res}
\scriptsize
\resizebox{0.95\linewidth}{!} 
{\begin{tabular}{c|rr|rrrr|rrrr}\toprule
\textbf{ID} &\textbf{\#ERP} &\textbf{\#AF} &\textbf{RT(s)} &\textbf{CFT(s)} &\textbf{TFT(s)} &\textbf{TotalT(s)} &\textbf{\#InTok} &\textbf{\#OutTok} &\textbf{\#P} &\textbf{Cost(\$)} \\\midrule
1 &8 &5 &188.81 &45.52 &10.65 &244.98 &46,068 &13,336 &39 &0.015 \\
2 &10 &9 &305.84 &123.32 &20.46 &449.61 &81,371 &27,257 &89 &0.029 \\
3 &6 &4 &747.44 &118.85 &215.55 &1,081.84 &222,652 &72,066 &94 &0.077 \\
4 &6 &5 &1,263.03 &161.50 &45.86 &1,470.39 &318,859 &57,769 &125 &0.082 \\
5 &22 &10 &1,182.65 &164.52 &11.70 &1,358.88 &268,129 &70,788 &146 &0.083 \\
6 &11 &10 &1,150.23 &264.11 &349.21 &1,763.55 &392,429 &87,263 &252 &0.111 \\
7 &17 &12 &1,595.18 &451.75 &66.10 &2,113.03 &427,183 &127,902 &187 &0.141 \\
8 &18 &15 &1,809.05 &537.59 &31.19 &2,377.83 &436,328 &123,892 &253 &0.140 \\
9 &17 &13 &2,930.73 &316.33 &436.54 &3,683.60 &928,716 &120,543 &278 &0.212 \\
10 &16 &11 &2,839.98 &330.11 &14.17 &3,184.26 &610,139 &112,067 &206 &0.159 \\
11 &15 &12 &3,365.71 &693.44 &158.15 &4,217.30 &1,102,037 &147,435 &322 &0.254 \\
12 &12 &12 &4,873.78 &491.10 &75.06 &5,439.94 &2,010,145 &160,203 &363 &0.398 \\
13 &38 &27 &4,743.65 &793.61 &238.83 &5,776.08 &1,504,725 &264,773 &511 &0.385 \\
14 &53 &43 &8,951.85 &2,658.11 &1,037.37 &12,647.33 &14,008,537 &418,288 &920 &2.352 \\
15 &80 &61 &8,293.31 &2,281.44 &862.65 &11,437.40 &9,699,548 &519,250 &1,137 &1.766 \\
16 &87 &70 &12,463.53 &650.76 &319.79 &13,434.08 &2,651,673 &234,328 &1,004 &0.538 \\
17 &60 &47 &6,510.92 &582.86 &586.44 &7,680.21 &579,508 &136,921 &531 &0.169 \\
18 &99 &80 &16,218.48 &4,631.28 &622.08 &21,471.84 &4,546,460 &952,967 &1,980 &1.254 \\
19 &73 &58 &18,867.08 &2,041.97 &1,963.26 &22,872.31 &4,831,868 &356,541 &1,490 &0.939 \\
20 &249 &132 &16,791.45 &2,422.30 &12,365.05 &31,578.80 &5,352,697 &881,196 &1,541 &1.332 \\
21 &157 &113 &20,378.32 &4,229.75 &2,410.53 &27,018.61 &8,878,129 &861,532 &1,882 &1.849 \\
22 &194 &158 &19,006.08 &8,199.58 &2,640.55 &29,846.22 &10,299,083 &960,924 &2,039 &2.121 \\
23 &105 &83 &12,430.74 &2,192.71 &8,888.03 &23,511.47 &60,813,477 &609,506 &1,594 &9.488 \\
24 &84 &68 &29,406.70 &1,595.97 &6,008.53 &37,011.20 &9,507,522 &442,569 &2,225 &1.692 \\
25 &158 &130 &16,422.75 &5,805.27 &6,840.01 &29,068.04 &2,625,267 &606,173 &2,128 &0.757 \\
26 &175 &152 &41,537.59 &3,229.05 &38,712.37 &83,479.01 &17,347,685 &662,482 &3,114 &3.000 \\
27 &138 &125 &36,488.11 &4,170.67 &20,440.27 &61,099.05 &3,322,335 &616,631 &2,570 &0.868 \\
28 &347 &257 &51,822.16 &5,775.61 &2,837.01 &60,434.78 &5,344,925 &973,554 &3,621 &1.386 \\
\bottomrule
\end{tabular}}
\vspace{-2mm}
\end{table*}

\section{Evaluation}
\label{sec:evaluation}

We implemented \toolname\ as a prototype with 1,072 lines of Rust and 1,187 lines of Python. It uses \texttt{tree-sitter} to extract raw pointers, struct definitions, and global variables from translated Rust code. Pointer lifting, statement refactoring, and repair phases are powered by \texttt{gpt-4o-mini}, chosen for its low cost (i.e., \$0.150/M input tokens and \$0.600/M output tokens~\cite{GPTpricing}), making it suitable for iterative transformation. To speed up compilation during error fixing, we use \texttt{cargo check}, which performs type and borrow checking without linking.
We allow up to five repair attempts per pointer (i.e., the value of $N$ in Section~\ref{sec:approach} is five) and set the temperature to 0.0 for greedy decoding.

\smallskip
We evaluate the effectiveness and efficiency of \toolname\ by answering the four research questions:
\begin{itemize}[leftmargin=*]
    \item \textbf{RQ1.} How effectively and efficiently does \toolname\ eliminate raw pointers in the Rust programs generated by \textsc{C2Rust}?
    \item \textbf{RQ2.} How does \toolname\ compare with existing techniques that transform C to safer Rust programs? 
    \item \textbf{RQ3.} Which kinds of data structures are introduced by \toolname\ to replace raw pointers?
    \item \textbf{RQ4.} How do the decision tree-based prompting and code change analysis contribute to the performance of \toolname?
\end{itemize}

%Please add the following packages if necessary:
%\usepackage{booktabs, multirow} % for borders and merged ranges
%\usepackage{soul}% for underlines
%\usepackage{xcolor,colortbl} % for cell colors
%\usepackage{changepage,threeparttable} % for wide tables
%If the table is too wide, replace \begin{table}[!htp]...\end{table} with
%\begin{adjustwidth}{-2.5 cm}{-2.5 cm}\centering\begin{threeparttable}[!htb]...\end{threeparttable}\end{adjustwidth}

% \cp{Raw pointer elimintation or rewriting?}
% \yf{I am not sure about difference between elimintation or rewriting, but we use rewrite in previous parts so I change to rewriting}

% \cp{@yifei, why you miss one project?}\yf{Benchmark 24(rgba) has no local raw pointers, id corrected}

\smallskip
\textbf{Benchmark.}
We adopt C projects from the evaluation set used in \textsc{Nopcrat}~\cite{hong2024don}, comprising widely used real-world applications also used in prior C-to-Rust studies~\cite{emre2021translating, emre2023aliasing}. Of the original 55 programs, 26 lack test cases and 1 contains raw pointers only in function parameters, which are outside our scope. We thus select the remaining 28 projects with test cases as our benchmark, which are listed by Table~\ref{tab:benchmark} in Appendix~\ref{appendix:appendix_bench}.
Overall, the benchmark spans programs from a few hundred to 136K LOC and 17 to 2,731 raw pointers, providing a diverse set for evaluating the effectiveness and scalability of \toolname.

\subsection{RQ1: Effectiveness and Efficiency}

\smallskip
\ \ \ \ \textbf{Setup and Metrics.}
To quantify the effectiveness of \toolname, we introduce the number of eliminated raw pointers as our main metric. Meanwhile, we collect the number of functions refactored during this process.
The eliminated raw pointers in these functions can improve the memory safety of the function usage.
To quantify the efficiency of \toolname, we measure the time cost, input/output token costs, prompting rounds, and financial cost. Notably, we do not include the time overhead introduced by \textsc{C2Rust}, as it is negligible compared to the overhead incurred during the rewriting and repair phases of \toolname.

\begin{comment}
. For effectiveness evaluation, we count raw pointers that are eliminated as our main metric, we also collect nuwmber of functions that are rewritten during this process. In terms of efficiency, we collect statistics of metrics about time, system resource and LLM API uses. Specifically for each project, we collect running time for every stage in our pipeline(we do not count time elapsed in process of translation from original C code to initial unsafe Rust code as they are available in their artifact), memory cost of translation, prompting rounds, input/output token costs and corresponding financial costs.
\end{comment}

% \cp{For each project, we collect the following statistics:}

% \begin{itemize}
%     \item The number of eliminated raw pointers.
%     \item The number of affected functions (containing the eliminated raw pointers)
%     \item Time cost of each different stage: C2Rust stage, rewrite stage, compilation, running test case
%     \item Memory cost
%     \item Prompting: Prompting rounds, input/output token costs
%     \item Financial cost
% \end{itemize}

%Please add the following packages if necessary:
%\usepackage{booktabs, multirow} % for borders and merged ranges
%\usepackage{soul}% for underlines
%\usepackage{xcolor,colortbl} % for cell colors
%\usepackage{changepage,threeparttable} % for wide tables
%If the table is too wide, replace \begin{table}[!htp]...\end{table} with
%\begin{adjustwidth}{-2.5 cm}{-2.5 cm}\centering\begin{threeparttable}[!htb]...\end{threeparttable}\end{adjustwidth}

\smallskip
\textbf{Results.} 
Table~\ref{tab:res} presents detailed evaluation results of \toolname. The column \textbf{\#ERP} shows the number of eliminated raw pointers. The column \textbf{\#AF} indicates the number of affected functions. On average, \toolname\ removes 18.57\% of raw pointers and transforms 13.05\% of functions into safer code. Notably, it eliminates 34.3\% of pointers in \texttt{grep-3.11} (ID = 20), demonstrating the effectiveness of \toolname\ when analyzing large projects.

The columns \textbf{RT}, \textbf{CFT}, and \textbf{TFT} represent time spent on raw pointer rewriting, compiler error fixing, and test case error fixing, respectively, with total time in \textbf{TotalT}. On average, \toolname\ takes 5.02 hours per project, with the longest taking~28 hours. The bottleneck in time is rewriting, averaging 74.26\% of total time.
Additionally, the columns \textbf{InTok}, \textbf{OutTok}, and \textbf{\#P} represent input/output tokens and prompt rounds during rewriting and repair. The column \textbf{Cost} reflects total prompting expense. Each successful rewrite requires 14 prompt rounds. Across all 28 benchmarks, the total cost is \$31.60, averaging just \$1.13 per project, highlighting the high efficiency of \toolname.

\subsection{RQ2: Comparison with Baselines}

\ \ \ \ \textbf{Setup and Metrics.} 
% \textsc{Laertes} conducts an empirical study on unsafe usages of Rust raw pointers, which demonstrates that a large portion of pointers are simply dereferenced, therefore, it is possible to make them safe by rewriting them to safe references with proper lifetime information. 
%\wcp{Only one baseline? Is it possible to add one more?}
We compare our approach with \textsc{Laertes}~\cite{emre2021translating}, an automated tool that aims to convert raw pointers into safe references. Specifically, \textsc{Laertes} applies an optimistic rewriting strategy and iteratively addresses lifetime inference and borrow-checking errors to achieve a successful compilation.
Another recent work, \textsc{Nopcrat}~\cite{hong2024don}, focuses on rewriting raw pointers in the parameters that propagate output values as side effects.
However, due to the difference in scope, the sets of raw pointers eliminated by \toolname\ and \textsc{Nopcrat} are disjoint thus we do not include it in our comparative evaluation.

%We run the artifact of \textsc{Laertes} provided by its authors upon the projects in our benchmark but observe stability issues in several programs. When \textsc{Laertes} encounters unrecoverable compiler errors, we terminate it early and report the number of the elimintated pointers eliminated up to that point.

% \xl{Check and we probably need to say why we only choose this one.}
%Due to the complexity of implementation, when encountering compiler errors that cannot be resolved by \textsc{Laertes}, we terminate the rewrite early. Consequently, if a program fails to compile, we consider this an upper bound on the number of raw pointers that the baseline can rewrite, as it relies on an optimistic initial transformation.

% \cp{Criteria of baseline selection. Briefly introduce how baselines work.}

% \cp{For each baseline, collect the following statistics:}

% \begin{itemize}
%     \item The number of eliminated raw pointers.
%     \item The number of the raw pointers that cannot be eliminated by existing baselines
%     \item Total time cost
%     \item Memory cost
% \end{itemize}

\smallskip
% \cp{In Project 12, why \toolname\ can uniquely elimintate 12 raw pointers while the total number of elimintated raw pointer is 10?}
% \cp{@yifei: I have suggested you using four sets indicating the elimintated raw pointers and using $S_1 \setminus S_2$ indicating the difference!!! Currently, Table 3 is very confusing.}

% \begin{figure*}[t]
% \centering
% \includegraphics[width=\linewidth]{Fig/caseRQ2.pdf}
% \caption{An example of rewriting a raw pointer with a slice and utilizing Rust inherent APIs}
% \label{fig:caseRQ2}
% \end{figure*}
% 
% 
% \cp{TODO: re-check the statistics in this paragraph as the content of Table is changed.}
% \yf{Updated}
\textbf{Results.} 
Table~\ref{tab:ablation} shows the comparison results of \toolname\ and the baselines. The column \textbf{ID} lists program identifiers (as in Table~\ref{tab:benchmark}). In the column \textbf{\textsc{Laertes}}, the sub-column $\boldsymbol{E_L}$ denotes the number of raw pointers eliminated by \textsc{Laertes}, while the sub-column $\boldsymbol{E_{L \setminus P}}$ indicates the number of the raw pointers \textsc{Laertes} eliminates and \toolname{} does not eliminate. The sub-column $T(s)$ indicates the execution time of \textsc{Laertes}. 
Similarly, the column \textbf{\toolname} includes the sub-column $\boldsymbol{E_P}$, which indicates the number of pointers eliminated by \toolname{}, and $\boldsymbol{E_{P \setminus L}}$, which indicates the number of raw pointers eliminated by \toolname{} and not eliminated by \textsc{Laertes}.
Overall, \toolname\ removes 2,255 raw pointers, compared to 506 by \textsc{Laertes}, and outperforms it on most projects.
The size of the project \texttt{buffer} (ID = 2) is small, enabling \textsc{Laertes} to infer lifetimes effectively and efficiently.
%For the other two projects, the test cases could not be executed correctly in our environment, which prevented data structure usage repair.
Notably, \toolname\ uniquely eliminates 2,160 raw pointers (95.79\% of its total), indicating that most rewrites are beyond the reach of the baseline \textsc{Laertes}. In contrast, \textsc{Laertes} only uniquely handles 438 raw pointers. Overall, \toolname\ eliminates 3.46× more raw pointers than \textsc{Laertes}, demonstrating its effectiveness.

In terms of efficiency, \textsc{Laertes} processes all 28 projects in 3,732.85 seconds (133.32 seconds per project), while \toolname\ averages 5.02 hours per project due to prompt-related overhead. However, with ongoing LLM performance improvements, \toolname\ is expected to become significantly faster over time.

\begin{figure}[t]
\centering
\includegraphics[width=0.6\linewidth]{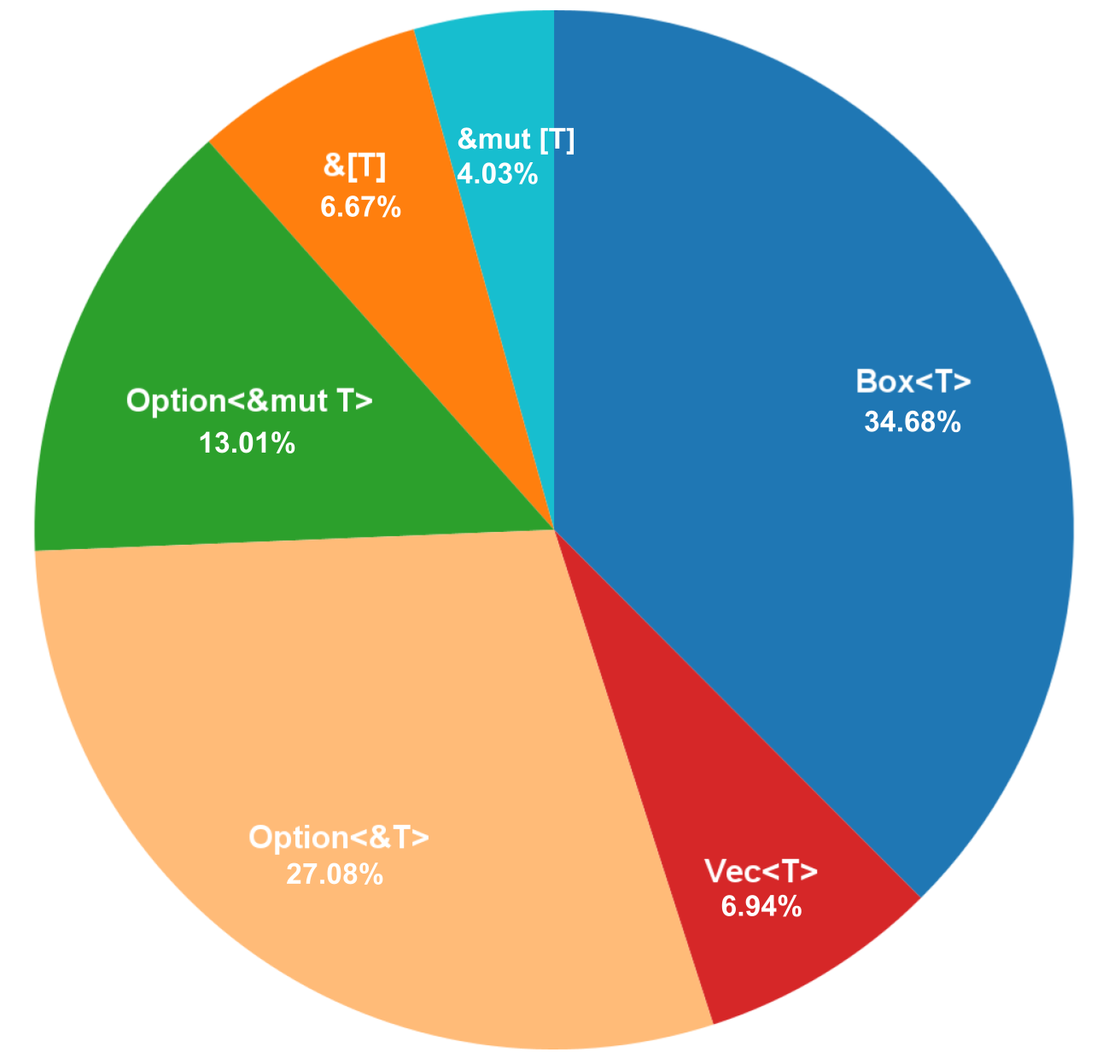}
\caption{The proportions of different data structures 
}
\vspace{-6mm}
\label{fig:catep}
\end{figure}

\subsection{RQ3: Categorization of Data Structures}
\ \ \textbf{Setup and Metrics.} To show the diversity of Rust data structures utilized by \toolname, we categorize Rust data structures used by \toolname\ to uniquely eliminate rewritten raw pointers. This analysis is important because, unlike \textsc{Laertes}, which primarily converts raw pointers to references, \toolname\ supports a broader spectrum of Rust data structures, thereby offering more improvements on memory safety.

% \cp{What does the last sentence mean? Why you only collect the raw pointers that are uniquely discovered by \toolname? Did you really collect the statistics in this way? In this section, we did not mention the baseline. Why do you emphasize the uniquely elimintated pointers in the absence of such context on baseline?}
% We list six categories which correspond to the six final state data structure in our decision tree.
% \cp{All the colomn names are shown in the form of the column \textbf{XXX}. Make them consistent with the ones in Section 6.1.}

\smallskip
\textbf{Results.} Figure~\ref{fig:catep} shows the distribution of Rust data structures used in pointer rewriting.
Overall, singleton data structures, e.g., \texttt{Box}, \texttt{Option<\&T>}, and \texttt{Option<\&mut T>}, account for the majority of rewritten raw pointers (74.8\%).
Meanwhile, a non-trivial portion (17.6\%) of raw pointers are successfully rewritten into sequential data structures, including \texttt{Vec}, \texttt{\&[T]}, and \texttt{\&mut [T]}.
In terms of ownership, 41.6\% of raw pointers are rewritten into \texttt{Box} and \texttt{Vec} which ensures that the lifetimes of the wrapped values are properly bounded,
thereby preventing memory leaks.
Lastly, 33.8\% of the raw pointers are converted into read-only structures (e.g., \texttt{Option<\&T>}, \texttt{\&[T]}), removing unnecessary mutability. We present a case demonstrating \toolname's capability of leveraging Rust inherent APIs.
We provide a case study of data structure usage within the generated Rust code in Appendix~\ref{appendix:case1}.

\begin{table}[t]\centering
\caption{The statistics of comparison with baseline \textsc{Laertes} and two ablations NT and NCC
%\wcp{Fix the data in ID 17 and ID 27.}
}
\vspace{-2mm}
\label{tab:ablation}
%\footnotesize
\resizebox{\linewidth}{!}{
\begin{tabular}{c|rrr|rr|rr|rr}
\toprule
\multirow{2}{*}{\textbf{Id}} &\multicolumn{3}{c}{\textbf{\textsc{Laertes}}} &\multicolumn{2}{c}{\textbf{\toolname}} &\multicolumn{2}{c}{\textbf{NT}} &\multicolumn{2}{c}{\textbf{NCC}}
\\\cmidrule{2-10}
 &$E_L$ &$E_{L \setminus P}$ &T(s) &$E_P$ &$E_{P \setminus L}$ &$E_{T}$ &$E_{T \setminus L}$ &$E_{C}$ &$E_{C \setminus L}$ \\\midrule
1 &2 &2 &1.15 &8 &8 &3 &3 &11 &10 \\
2 &27 &17 &2.41 &10 &0 &8 &2 &12 &1 \\
3 &0 &0 &1.74 &6 &6 &1 &1 &2 &2 \\
4 &5 &2 &1.44 &6 &3 &5 &3 &4 &2 \\
5 &3 &2 &4.35 &22 &21 &7 &7 &10 &10 \\
6 &1 &0 &5.64 &11 &10 &7 &7 &7 &7 \\
7 &0 &0 &1.73 &17 &17 &6 &6 &9 &9 \\
8 &4 &2 &4.11 &18 &15 &11 &9 &16 &13 \\
9 &2 &2 &7.11 &17 &17 &7 &6 &0 &0 \\
10 &12 &11 &7.74 &16 &15 &8 &6 &8 &7 \\
11 &3 &3 &25.55 &15 &15 &7 &7 &7 &7 \\
12 &14 &12 &5.54 &12 &10 &2 &2 &4 &3 \\
13 &5 &3 &4.25 &38 &35 &8 &7 &15 &15 \\
14 &9 &6 &121.46 &53 &48 &36 &33 &40 &35 \\
15 &14 &12 &46.22 &80 &77 &45 &41 &61 &56 \\
16 &7 &7 &15.08 &87 &86 &56 &55 &68 &67 \\
17 &12 &12 &42.42 &60 &58 &54 &52 &62 &60 \\
18 &2 &1 &90.6 &99 &98 &0 &0 &0 &0 \\
19 &28 &28 &18.2 &73 &73 &91 &85 &71 &69 \\
20 &52 &39 &45.38 &249 &227 &80 &75 &68 &65 \\
21 &42 &40 &47.66 &157 &155 &96 &95 &95 &93 \\
22 &43 &36 &95.03 &194 &187 &108 &102 &124 &118 \\
23 &37 &24 &35.06 &105 &88 &0 &0 &0 &0 \\
24 &10 &10 &74.99 &84 &84 &0 &0 &41 &41 \\
25 &58 &53 &713.13 &158 &153 &142 &132 &177 &168 \\
26 &37 &37 &164.86 &175 &170 &187 &179 &169 &164 \\
27 &40 &40 &1619.28 &138 &138 &95 &94 &106 &104 \\
28 &37 &37 &530.72 &347 &346 &138 &138 &393 &392 \\
\bottomrule
\end{tabular}
}
\vspace{-7mm}
\end{table}

\subsection{RQ4: Ablation Study}
\label{subsec:ablation}
\ \ \ \ \textbf{Setup and Metrics.} To quantify how the technical designs of \toolname\ contribute to the rewriting results, we introduce two ablations: \textsc{NT} and \textsc{NCC}, which disable decision tree-based prompting and code change analysis in the pointer lifting and data structure usage repair, respectively.
Specifically, the ablation \textsc{NT} enforces the LLM to directly choose a data structure from the six data structures shown in the leaf nodes in Figure~\ref{fig:tree}, without conducting semantic analyses. The ablation \textsc{NCC} feeds the functions before and after raw pointer rewriting to the LLM along with error messages, but without any code change information.
We use the number of raw pointers eliminated under different ablations as the primary evaluation metric.

% \cp{Introduce two ablations}

% \cp{The metrics should be the same as the ones in RQ1}
% \xl{The result text is inconsistent with the table.}

\smallskip
\textbf{Results.} Table~\ref{tab:ablation} reports results for ablations \textsc{NT} and \textsc{NCC}. Under the column \textbf{NT}, the sub-columns $\boldsymbol{E_T}$ and $\boldsymbol{E_{T \setminus L}}$ denote the total and uniquely eliminated pointers, respectively; \textbf{NCC} shows the same for its setting. As shown, \textsc{NT} eliminates 1,208 pointers (46.43\% fewer than the full \toolname), and \textsc{NCC} eliminates 1,580 (29.93\% fewer). Removing the decision tree prompt causes a 46.43\% drop in unique eliminations; omitting code change analysis leads to a 29.93\% drop.
These results confirm the importance of both components—especially the decision tree—in enhancing rewrite quality. Even in reduced forms, both variants outperform \textsc{Laertes}, highlighting the robustness of LLM-driven rewriting.
Appendix~\ref{appendix:case2} provides a case study that shows an example handled by full-featured \toolname{} but not addressed by its ablation NT.

\subsection{Discussion}
\ \ \ \textbf{User Study.}
To assess semantic correctness, we recruited ten Rust-experienced volunteers to review outputs from \toolname. Each independently evaluated 20 randomly selected raw pointer eliminations (around 1\% of all cases), using \texttt{gpt-4o-mini} explanations and Rust documentation to judge semantic equivalence. Reviews took about 45 minutes each, and all volunteers confirmed the rewritten code preserved original semantics. 
These results suggest \toolname\ is a practical tool for C-to-Rust translation. While it lacks formal guarantees, it achieves strong empirical outcomes. With modest human effort, users can semi-automatically produce safer, idiomatic Rust code.

\smallskip
\textbf{Threats to Validity.}
The primary threat to internal validity stems from the stochastic nature of LLMs, which can yield nondeterministic outputs and affect reproducibility. As noted in Section~\ref{subsec:ablation}, two ablation settings slightly outperform the full-featured \toolname\ on \texttt{tar-1.34} and \texttt{glpk-5.0}. To address this, we fix the temperature to 0 for deterministic behavior.
External validity depends on test case quality, which influences both the elimination number and correctness. Our test-driven approach reduces manual effort, and the user study confirms that \toolname\ preserves semantic equivalence. Since broader test coverage could further improve rewriting quality, we calculated the coverage of test suites in our benchmark projects, observing 62.84\% on average. We then focused on rewritten pointers in functions not covered by existing test suites and leveraged LLMs to synthesize test cases and run them, in 59 total cases we found only 2 incorrect pointer rewrites, corresponding to a 96.61\% correctness rate among inspected cases, which further supports the correctness of our transformation pipeline.
Construct validity relates to LLM choice. We use \texttt{gpt-4o-mini}, a cost-effective and widely available model, and validate our findings with \texttt{Claude-3.5-Sonnet} and \texttt{Deepseek-v3.2} on the top 10 projects in Table~\ref{tab:res} with results demonstrated in Table~\ref{tab:additional}, where $E_{Cl}$,  $E_{Dp}$ correspond to pointers eliminated by \texttt{Claude-3.5-Sonnet} and \texttt{Deepseek-v3.2}, and $E_{Cl} \setminus E_{l}$, $E_{Dp} \setminus E_{l}$ correspond to pointers uniquely eliminated by \texttt{Claude-3.5-Sonnet} and \texttt{Deepseek-v3.2}. Compared with \texttt{gpt-4o-mini}, \texttt{Claude-3.5-Sonnet} and \texttt{Deepseek-v3.2} demonstrate similar performance in most projects and significantly better performance in some projects (1, 2, 8, 10). As LLMs advance and inference costs drop, the effectiveness and efficiency of \toolname{} will continue to improve.

% \begin{table}[h] \centering \small \begin{tabular}{c|cc} \hline Proj & Elim & Unique \\ \hline 1 & 14 & 12 \\ 2 & 11 & 1 \\ 3 & 4 & 4 \\ 4 & 6 & 5 \\ 5 & 24 & 23 \\ 6 & 20 & 19 \\ 7 & 18 & 18 \\ 8 & 42 & 41 \\ 9 & 22 & 22 \\ 10 & 32 & 31 \\ \hline \end{tabular} \caption{Claude 3.5 Sonnet Pointer Elimination Results} \label{tab:claude35} \end{table} \begin{table}[h] \centering \small \begin{tabular}{c|cc} \hline Proj & Elim & Unique \\ \hline 1 & 14 & 12 \\ 2 & 29 & 1 \\ 3 & 10 & 10 \\ 4 & 6 & 5 \\ 5 & 20 & 17 \\ 6 & 24 & 23 \\ 7 & 18 & 18 \\ 8 & 34 & 33 \\ 9 & 27 & 26 \\ 10 & 21 & 18 \\ \hline \end{tabular} \caption{DeepSeek v3.2 rPointer Elimination Results} \label{tab:deepseek} \end{table}
\begin{table}[h] \centering \small \begin{tabular}{c|cccc} \hline \textbf{Id} & $E_{Cl}$ & $E_{Cl} \setminus E_{L}$ & $E_{Dp}$ & $E_{Dp} \setminus E_{L}$ \\ \hline 1 & 14 & 12 & 14 & 12 \\ 2 & 11 & 1 & 29 & 1 \\ 3 & 4 & 4 & 10 & 10 \\ 4 & 6 & 5 & 6 & 5 \\ 5 & 24 & 23 & 20 & 17 \\ 6 & 20 & 19 & 24 & 23 \\ 7 & 18 & 18 & 18 & 18 \\ 8 & 42 & 41 & 34 & 33 \\ 9 & 22 & 22 & 27 & 26 \\ 10 & 32 & 31 & 21 & 18 \\ \hline \end{tabular} \caption{Additional LLM Pointer Elimination Results} \label{tab:additional} \end{table}
% Start of old version
% The primary threat to internal validity lies in the stochastic nature of the LLMs, which can produce nondeterministic outputs, posing challenges to reproducibility and consistency. For instance, as observed in Section~\ref{subsec:ablation}, the two ablation settings slightly outperform the full-featured \toolname\ on the projects \texttt{tar-1.34} and \texttt{glpk-5.0}. To mitigate such issue, we set the temperature to 0 in our implementation to enforce deterministic behavior.
% External validity is influenced by the quality of the test cases, which affects both the number of raw pointers eliminated and the correctness of the rewriting. Our test-driven approach significantly reduces manual effort in C-to-Rust translation, and the user study empirically demonstrates that \toolname\ preserves functional equivalence. More comprehensive test suites with higher coverage could further enhance the quality of the rewriting.
% Construct validity is affected by the choice of LLM, which may impact the transformation results. In our implementation, we adopt \textsf{gpt-4o-mini}, a cost-effective and widely available model. We also conduct a controlled experiment using \textsf{Claude-3.5-Sonnet} on the top 10 projects in Table~\ref{tab:res}, and observe that the results closely match those produced by \toolname\ powered by \textsf{gpt-4o-mini}. As LLM capabilities continue to improve and inference costs decline, both the effectiveness and efficiency of \toolname\ will improve.
\section{Conclusion}
This paper presents \toolname, a novel approach for transforming C programs into safer Rust programs by eliminating raw pointers using LLMs. Building on syntax-guided C-to-Rust translation, we introduce a raw pointer rewriting technique that targets raw pointers within single functions and replaces them with appropriate Rust data structures. Specifically, \toolname\ employs LLMs to lift raw pointers to Rust data structures and iteratively repairs the transformed program by focusing on the changed code.
Our evaluation on 28 real-world C programs demonstrates that \toolname, powered by \texttt{gpt-4o-mini}, eliminates a total of 2,255 raw pointers, with an average processing time of 5.02 hours and a cost of \$1.13 per project. We hope that \toolname\ provides valuable insights for advancing program translation techniques aimed at improving system reliability.

\section{Limitations}
Although \toolname\ effectively refactors specific raw pointers using Rust data structures, it has several drawbacks. First, it performs raw pointer rewriting within individual functions, keeping type signatures unchanged, as shown in Section~\ref{subsubsec:pr}. While this conservative scope simplifies repairs, it misses chances to eliminate raw-pointer parameters, unlike \textsc{Nopcrat}~\cite{hong2024don}. Future work could enable more expressive rewriting by modifying function signatures across call chains. Second, \toolname\ lacks a theoretical correctness guarantee; without strong or sufficient test coverage, raw pointer elimination can break functional equivalence despite passing compilation and tests. Enhancements could involve synthesizing unit tests using LLMs for differential testing~\cite{zhou2025lwdiff,zhang2025drwasi, ParCleanse, rao2024diffspec} and applying verification techniques like relational verification~\cite{DBLP:conf/cav/UnnoTK21,DBLP:conf/pldi/ChurchillP0A19,Pardiff}. Finally, the rewritten code may have low readability and maintainability, since current efforts focus mainly on raw pointer elimination. We plan to use LLMs in a post-processing phase to further refine and improve the quality of the refactored Rust code.

% \section*{Data Availability}
% The artifact that supports Section~\ref{sec:approach}, Section~\ref{sec:implementation}, and Section~\ref{sec:evaluation} has been uploaded to an anonymous repository\footnote{\url{https://anonymous.4open.science/r/PR2-7FA9}}. We will make \toolname\ publicly available upon publication.

\section*{Acknowledgement}
We are grateful to the Center for AI Safety for providing computational resources. This work
was funded in part by the National Science Foundation Awards SHF-1901242, SHF-1910300,
Proto-OKN 2333736, IIS2416835, DARPA VSPELLS - HR001120S0058, ONR N00014-23-1-2081, and
Amazon. Any opinions, findings and conclusions or recommendations expressed in this material
are those of the authors and do not necessarily reflect the views of the sponsors.

\bibliography{bib/ref}

@String{Computing = "Computing" }

@String{Computer = "{IEEE} Computer" }

@String{Academic = "Academic Press" }

@String{Springer = "Springer-Verlag" }

@article{yuan2024transagent,
  title={Transagent: An llm-based multi-agent system for code translation},
  author={Yuan, Zhiqiang and Chen, Weitong and Wang, Hanlin and Yu, Kai and Peng, Xin and Lou, Yiling},
  journal={arXiv preprint arXiv:2409.19894},
  year={2024}
}

@article{yang2024exploring,
  title={Exploring and unleashing the power of large language models in automated code translation},
  author={Yang, Zhen and Liu, Fang and Yu, Zhongxing and Keung, Jacky Wai and Li, Jia and Liu, Shuo and Hong, Yifan and Ma, Xiaoxue and Jin, Zhi and Li, Ge},
  journal={Proceedings of the ACM on Software Engineering},
  volume={1},
  number={FSE},
  pages={1585--1608},
  year={2024},
  publisher={ACM New York, NY, USA},
url = {https://doi.org/10.1145/3660778}
}

@article{hong2025type,
  title={Type-migrating C-to-Rust translation using a large language model},
  author={Hong, Jaemin and Ryu, Sukyoung},
  journal={Empirical Software Engineering},
  volume={30},
  number={1},
  pages={3},
  year={2025},
  publisher={Springer},
  url = {https://doi.org/10.1007/s10664-024-10573-2}
}

@inproceedings{DBLP:conf/cav/UnnoTK21,
  author       = {Hiroshi Unno and
                  Tachio Terauchi and
                  Eric Koskinen},
  editor       = {Alexandra Silva and
                  K. Rustan M. Leino},
  title        = {Constraint-Based Relational Verification},
  booktitle    = {Computer Aided Verification - 33rd International Conference, {CAV}
                  2021, Virtual Event, July 20-23, 2021, Proceedings, Part {I}},
  series       = {Lecture Notes in Computer Science},
  volume       = {12759},
  pages        = {742--766},
  publisher    = {Springer},
  year         = {2021},
  url          = {https://doi.org/10.1007/978-3-030-81685-8\_35},
  doi          = {10.1007/978-3-030-81685-8\_35},
  bibsource    = {dblp computer science bibliography, https://dblp.org}
}

@inproceedings{DBLP:conf/pldi/ChurchillP0A19,
  author       = {Berkeley R. Churchill and
                  Oded Padon and
                  Rahul Sharma and
                  Alex Aiken},
  editor       = {Kathryn S. McKinley and
                  Kathleen Fisher},
  title        = {Semantic program alignment for equivalence checking},
  booktitle    = {Proceedings of the 40th {ACM} {SIGPLAN} Conference on Programming
                  Language Design and Implementation, {PLDI} 2019, Phoenix, AZ, USA,
                  June 22-26, 2019},
  pages        = {1027--1040},
  publisher    = {{ACM}},
  year         = {2019},
  url          = {https://doi.org/10.1145/3314221.3314596},
  doi          = {10.1145/3314221.3314596},
  bibsource    = {dblp computer science bibliography, https://dblp.org}
}

@article{hong2024don,
  title={Don’t Write, but Return: Replacing Output Parameters with Algebraic Data Types in C-to-Rust Translation},
  author={Hong, Jaemin and Ryu, Sukyoung},
  journal={Proceedings of the ACM on Programming Languages},
  volume={8},
  number={PLDI},
  pages={716--740},
  year={2024},
  publisher={ACM New York, NY, USA}
}

@inproceedings{zhang2023ownership,
  title={Ownership guided C to Rust translation},
  author={Zhang, Hanliang and David, Cristina and Yu, Yijun and Wang, Meng},
  booktitle={International Conference on Computer Aided Verification},
  pages={459--482},
  year={2023},
  organization={Springer}
}

@misc{GPTpricing,
    title = {API Pricing},
    url = {https://openai.com/api/pricing/},
    lastaccessed = {February 12, 2025},
    year =         {2025},
    author = "OpenAI"
}

@article{emre2021translating,
  title={Translating C to safer Rust},
  author={Emre, Mehmet and Schroeder, Ryan and Dewey, Kyle and Hardekopf, Ben},
  journal={Proceedings of the ACM on Programming Languages},
  volume={5},
  number={OOPSLA},
  pages={1--29},
  year={2021},
  publisher={ACM New York, NY, USA}
}

@article{emre2023aliasing,
  title={Aliasing limits on translating C to safe Rust},
  author={Emre, Mehmet and Boyland, Peter and Parekh, Aesha and Schroeder, Ryan and Dewey, Kyle and Hardekopf, Ben},
  journal={Proceedings of the ACM on Programming Languages},
  volume={7},
  number={OOPSLA1},
  pages={551--579},
  year={2023},
  publisher={ACM New York, NY, USA}
}

@misc{c2rust,
    title = "C2rust Demonstration",
    url = "https://c2rust.com/",
    lastaccessed = "February 24, 2025",
    year =         "2025",
    author = "Galois and Immunant"
}

@misc{citrus,
    title = {Citrus: Convert C to Rust},
    url = {https://gitlab.com/citrus-rs/citrus#citrus-convert-c-to-rust},
    lastaccessed = {March 15, 2025},
    year =         {2025},
    author = "Citrus-rs"
}

@misc{corrode,
    title = "Corrode: Automatic semantics-preserving
translation from c to rust",
    url = "https://github.com/jameysharp/corrode",
    lastaccessed = "March 4, 2025",
    year =         "2025",
    author = "Jamey Sharp"
}

@article{shetty2024syzygy,
  title={Syzygy: Dual Code-Test C to (safe) Rust Translation using LLMs and Dynamic Analysis},
  author={Shetty, Manish and Jain, Naman and Godbole, Adwait and Seshia, Sanjit A and Sen, Koushik},
  journal={arXiv preprint arXiv:2412.14234},
  year={2024}
}

@article{yang2024vert,
  title={VERT: Verified equivalent rust transpilation with large language models as few-shot learners},
  author={Yang, Aidan ZH and Takashima, Yoshiki and Paulsen, Brandon and Dodds, Josiah and Kroening, Daniel},
  journal={arXiv preprint arXiv:2404.18852},
  year={2024}
}

@article{yin2024rectifier,
  title={Rectifier: Code translation with corrector via llms},
  author={Yin, Xin and Ni, Chao and Nguyen, Tien N and Wang, Shaohua and Yang, Xiaohu},
  journal={arXiv preprint arXiv:2407.07472},
  year={2024}
}

@inproceedings{cui2024code,
  title={Code Comprehension: Review and Large Language Models Exploration},
  author={Cui, Jielun and Zhao, Yutong and Yu, Chong and Huang, Jiaqi and Wu, Yuanyuan and Zhao, Yu},
  booktitle={2024 IEEE 4th International Conference on Software Engineering and Artificial Intelligence (SEAI)},
  pages={183--187},
  year={2024},
  organization={IEEE}
}

@inproceedings{nam2024using,
  title={Using an llm to help with code understanding},
  author={Nam, Daye and Macvean, Andrew and Hellendoorn, Vincent and Vasilescu, Bogdan and Myers, Brad},
  booktitle={Proceedings of the IEEE/ACM 46th International Conference on Software Engineering},
  pages={1--13},
  year={2024}
}

@article{bouzenia2024repairagent,
  title={Repairagent: An autonomous, llm-based agent for program repair},
  author={Bouzenia, Islem and Devanbu, Premkumar and Pradel, Michael},
  journal={arXiv preprint arXiv:2403.17134},
  year={2024}
}

@inproceedings{kulsum2024case,
  title={A case study of llm for automated vulnerability repair: Assessing impact of reasoning and patch validation feedback},
  author={Kulsum, Ummay and Zhu, Haotian and Xu, Bowen and d'Amorim, Marcelo},
  booktitle={Proceedings of the 1st ACM International Conference on AI-Powered Software},
  pages={103--111},
  year={2024}
}

@inproceedings{10.1145/3650212.3680323,
author = {Xia, Chunqiu Steven and Zhang, Lingming},
title = {Automated Program Repair via Conversation: Fixing 162 out of 337 Bugs for \$0.42 Each using ChatGPT},
year = {2024},
isbn = {9798400706127},
publisher = {Association for Computing Machinery},
address = {New York, NY, USA},
url = {https://doi.org/10.1145/3650212.3680323},
doi = {10.1145/3650212.3680323},
booktitle = {Proceedings of the 33rd ACM SIGSOFT International Symposium on Software Testing and Analysis},
pages = {819–831},
numpages = {13},
location = {Vienna, Austria},
series = {ISSTA 2024}
}

@incollection{jana2024cotran,
  title={Cotran: An llm-based code translator using reinforcement learning with feedback from compiler and symbolic execution},
  author={Jana, Prithwish and Jha, Piyush and Ju, Haoyang and Kishore, Gautham and Mahajan, Aryan and Ganesh, Vijay},
  booktitle={ECAI 2024},
  pages={4011--4018},
  year={2024},
  publisher={IOS Press}
}

@inproceedings{zhou2025lwdiff,
  title={LWDIFF: An LLM-Assisted Differential Testing Framework for WebAssembly Runtimes},
  author={Zhou, Shiyao and Wang, Jincheng and Ye, He and Zhou, Hao and Le Goues, Claire and Luo, Xiapu},
  booktitle={2025 IEEE/ACM 47th International Conference on Software Engineering (ICSE)},
  pages={769--769},
  year={2025},
  organization={IEEE Computer Society}
}

@article{zhang2025drwasi,
  title={DrWASI: LLM-assisted Differential Testing for WebAssembly System Interface Implementations},
  author={Zhang, Yixuan and He, Ningyu and Gao, Jianting and Cao, Shangtong and Liu, Kaibo and Wang, Haoyu and Ma, Yun and Huang, Gang and Liu, Xuanzhe},
  journal={ACM Transactions on Software Engineering and Methodology},
  year={2025},
  publisher={ACM New York, NY}
}

@article{rao2024diffspec,
  title={DiffSpec: Differential Testing with LLMs using Natural Language Specifications and Code Artifacts},
  author={Rao, Nikitha and Gilbert, Elizabeth and Ramananandro, Tahina and Swamy, Nikhil and Goues, Claire Le and Fakhoury, Sarah},
  journal={arXiv preprint arXiv:2410.04249},
  year={2024}
}

@article{van2021toward,
  title={Toward unseating the unsafe C programming language},
  author={van Oorschot, Paul C},
  journal={IEEE Security \& Privacy},
  volume={19},
  number={02},
  pages={4--6},
  year={2021},
  publisher={IEEE Computer Society}
}

@article{van2023memory,
  title={Memory errors and memory safety: C as a case study},
  author={van Oorschot, Paul C},
  journal={IEEE Security \& Privacy},
  volume={21},
  number={2},
  pages={70--76},
  year={2023},
  publisher={IEEE}
}

@article{lord2023urgent,
  title={The urgent need for memory safety in software products},
  author={Lord, Bob},
  journal={Cybersecurity \& Infrastructure Security Agency},
  year={2023}
}

@inproceedings{yang2018source,
  title={The source and exploitation of the program vulnerability},
  author={Yang, Gao and others},
  booktitle={2018 3rd Joint International Information Technology, Mechanical and Electronic Engineering Conference (JIMEC 2018)},
  pages={89--94},
  year={2018},
  organization={Atlantis Press}
}

@article{turner2014security,
  title={Security vulnerabilities of the top ten programming languages: C, Java, C++, Objective-C, C\#, PHP, Visual Basic, Python, Perl, and Ruby},
  author={Turner, Stephen},
  journal={Journal of Technology Research},
  volume={5},
  pages={1},
  year={2014},
  publisher={Academic and Business Research Institute (AABRI)}
}

@inproceedings{avots2005improving,
  title={Improving software security with a C pointer analysis},
  author={Avots, Dzintars and Dalton, Michael and Livshits, V Benjamin and Lam, Monica S},
  booktitle={Proceedings of the 27th international conference on Software engineering},
  pages={332--341},
  year={2005}
}

@inproceedings{matsakis2014rust,
  title={The rust language},
  author={Matsakis, Nicholas D and Klock, Felix S},
  booktitle={Proceedings of the 2014 ACM SIGAda annual conference on High integrity language technology},
  pages={103--104},
  year={2014}
}

@article{zheng2025large,
  title={Large Language Models for Validating Network Protocol Parsers},
  author={Zheng, Mingwei and Xie, Danning and Zhang, Xiangyu},
  journal={arXiv preprint arXiv:2504.13515},
  year={2025}
}

@inproceedings{ParCleanse,
    author = {Zheng, Mingwei and Xie, Danning and Shi, Qingkai and Wang, Chengpeng and Zhang, Xiangyu},
    title = {Validating Network Protocol Parsers with Traceable RFC Document Interpretation},
    booktitle = {Proceedings of the 34th ACM SIGSOFT International Symposium on Software Testing and Analysis},
    series = {ISSTA 2025}
}

@inproceedings{Pardiff,
    author = {Zheng, Mingwei and Shi, Qingkai and Liu, Xuwei and Xu, Xiangzhe and Yu, Le and Liu, Congyu and Wei, Guannan and Zhang, Xiangyu},
    title = {ParDiff: Practical Static Differential Analysis of Network Protocol Parsers}, 
    booktitle = {Proc. ACM Program. Lang.},
    publisher = {ACM},
    pages        = {1208--1234},
    year = {2024},
    series = {OOPSLA '24},
    doi = {10.1145/3649854},
}
\appendix
\newpage
\section{Appendix}
\label{sec:appendix}

\begin{table}[t]\centering
\caption{The statistics of benchmark programs}\label{tab:benchmark}
\vspace{-2mm}
\resizebox{0.98\linewidth}{!}{
\begin{tabular}{cl|rrrr}\toprule
\textbf{ID} &\textbf{Project Name} &\textbf{C Size} &\textbf{Rust Size} &\textbf{\#F} &\textbf{\#RP} \\\midrule
1 &quadtree &437 &1,180 &31 &17 \\
2 &buffer &452 &1,111 &40 &30 \\
3 &genann &690 &2,325 &27 &30 \\
4 &libtree &1,412 &2,617 &32 &41 \\
5 &urlparser &440 &1,380 &21 &52 \\
6 &ed-1.19 &2,584 &5,462 &137 &80 \\
7 &mcsim-6.2.0 &20,033 &35,558 &506 &80 \\
8 &hello-2.12.1 &37,192 &10,372 &176 &88 \\
9 &indent-2.2.13 &20,927 &15,066 &128 &100 \\
10 &pth-2.0.7 &8,797 &12,649 &235 &102 \\
11 &gzip-1.12 &53,751 &21,463 &244 &122 \\
12 &pexec-1.0rc8 &5,684 &12,027 &158 &156 \\
13 &units-2.22 &7,253 &11,188 &144 &185 \\
14 &patch-2.7.6 &54,350 &114,076 &601 &361 \\
15 &cpio-2.14 &71,819 &84,012 &707 &431 \\
16 &cflow-1.7 &45,223 &25,467 &480 &476 \\
17 &enscript-1.6.6 &38,424 &78,159 &259 &523 \\
18 &libosip2-5.3.1 &18,848 &34,195 &726 &628 \\
19 &rcs-5.10.1 &63,254 &35,811 &496 &664 \\
20 &grep-3.11 &117,416 &84,706 &867 &725 \\
21 &diffutils-3.10 &112,943 &96,418 &853 &760 \\
22 &findutils-4.9.0 &136,421 &143,361 &1,232 &835 \\
23 &sed-4.9 &91,818 &66,580 &675 &835 \\
24 &make-4.4.1 &31,866 &35,395 &445 &997 \\
25 &wget-1.21.4 &\textbf{136,566} &\textbf{189,761} &1,355 &1,320 \\
26 &tar-1.34 &102,864 &137,655 &\textbf{1,669} &1,399 \\
27 &gawk-5.2.2 &65,617 &134,098 &1,450 &1,792 \\
28 &glpk-5.0 &77,420 &138,372 &1,523 &\textbf{2,731} \\
\bottomrule
\end{tabular}
}
\vspace{-2mm}
\end{table}

% \begin{figure}[t]
% 	\centering
% 	\includegraphics[width=\linewidth]{Fig/errfix2-new.pdf}
%     \vspace{-6mm}
% 	\caption{An example of testcase error fixing
%     }
% 	\label{fig:errfix2}
% 	\vspace{-3mm}
% \end{figure}

\begin{figure}[t]
	\centering
	\includegraphics[width=\linewidth]{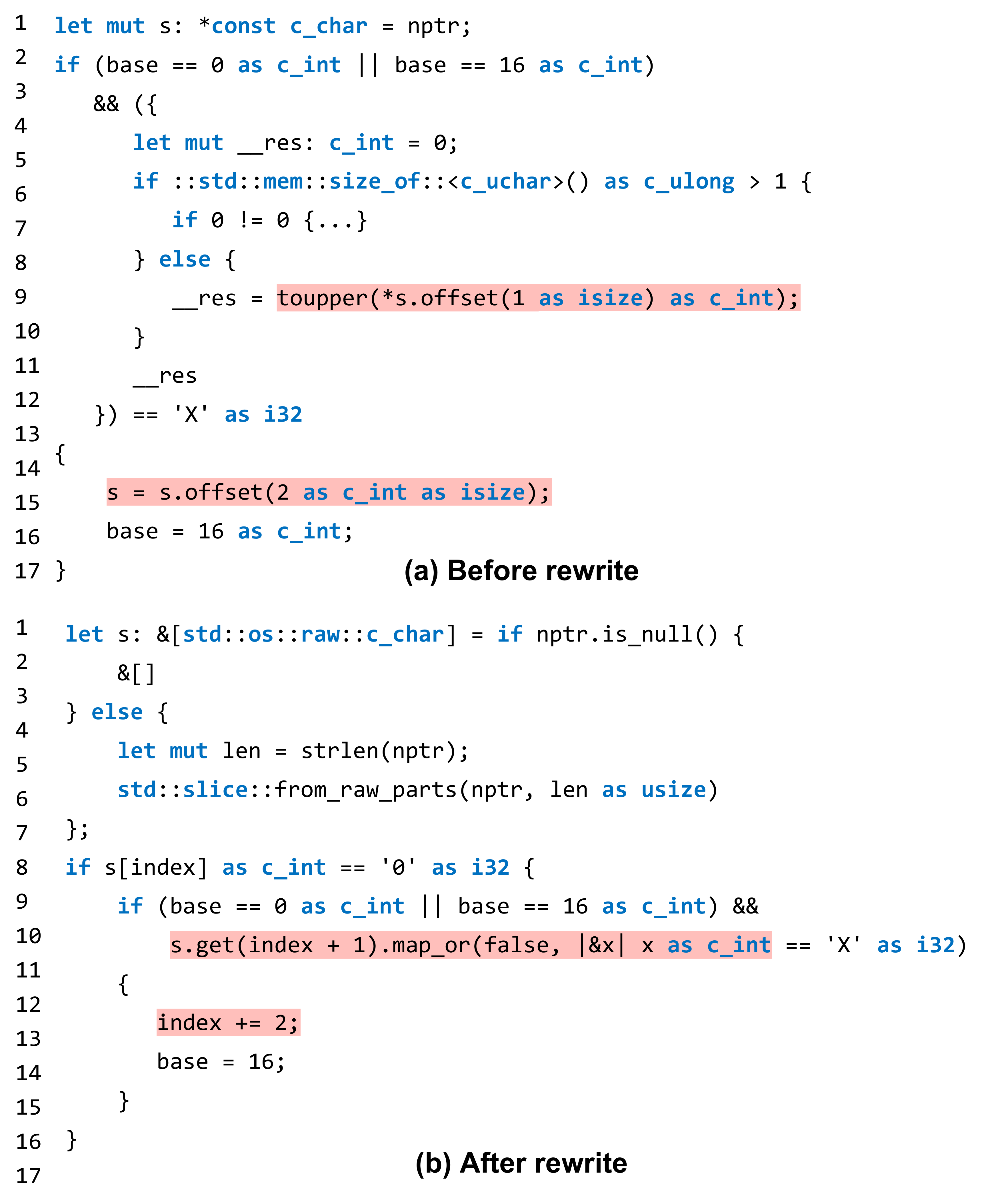}
    \vspace{-6mm}
	\caption{An example of rewriting a raw pointer with a slice and utilizing Rust inherent APIs}
	\label{fig:caseRQ2}
	\vspace{-5mm}
\end{figure}

\subsection{Details of Experimental Subjects}
\label{appendix:appendix_bench}
The details of the experimental subjects are shown in Table~\ref{tab:benchmark}.
Specifically, the columns \textbf{C Size} and \textbf{Rust Size} report the sizes of the original and \textsc{C2Rust}-translated code, respectively.
The columns \textbf{\#F} and \textbf{\#RP} show the number of functions and raw pointers in the Rust output, respectively.
The benchmark spans programs from a few hundred to 136K LOC and 17 to 2,731 raw pointers, providing a diverse set for evaluating \toolname’s effectiveness and scalability.

% \subsection{An Example of Testcase Error Fixing}
% \label{appendix:testcase_errror_fixing}
% Figure~\ref{fig:errfix2} presents a fix for a test case failure caused by out-of-bounds vector access by using the API \texttt{push} guided by the code change analysis.
% We only provide the result of the code change analysis to narrow down the buggy snippets, which facilitates the LLMs to fix the test case error more effectively.

\subsection{An Example of Rust Data Structures}
\label{appendix:case1}
We present a case study demonstrating how \toolname\ effectively transforms C code into safer and more idiomatic Rust by leveraging Rust’s data structures and APIs.
As shown in Figure~\ref{fig:caseRQ2}, the code snippet is derived from a variant of the C standard library function \textit{strtoull}, which parses C strings into integer values based on a specified numeric base. Specifically, the original Rust program generated by the tool \textsc{C2Rust}, which is shown in Figure~\ref{fig:caseRQ2}(a), contains unreachable branches caused by the unsatisfiable conditions at lines 5 and 7. Meanwhile, it relies on raw pointer arithmetic via \texttt{offset}, further introducing potential invalid memory operations.
After the raw pointer rewriting and data structure usage repair, \toolname\ replaces the raw pointer \texttt{s} with a Rust slice (declared and initialized at line 5) and utilizes the API \texttt{get} to safely access elements by index, returning an \texttt{Option} instead of performing unchecked access. \toolname\ also employs the API \texttt{map\_or}, which allows the program to apply a closure to the retrieved character if it exists, or return a default value \texttt{false} when the index is out of bounds. This effectively eliminates the need for fragile conditional logic and pointer arithmetic.
This case illustrates the empirical benefit of \toolname\ in semantically understanding the original C logic and intelligently selecting Rust data structures as the abstractions of raw pointers through pointer lifting.

\subsection{An Example of Ablation Study}
\label{appendix:case2}

\begin{figure}[t]
	\centering
	\includegraphics[width=\linewidth]{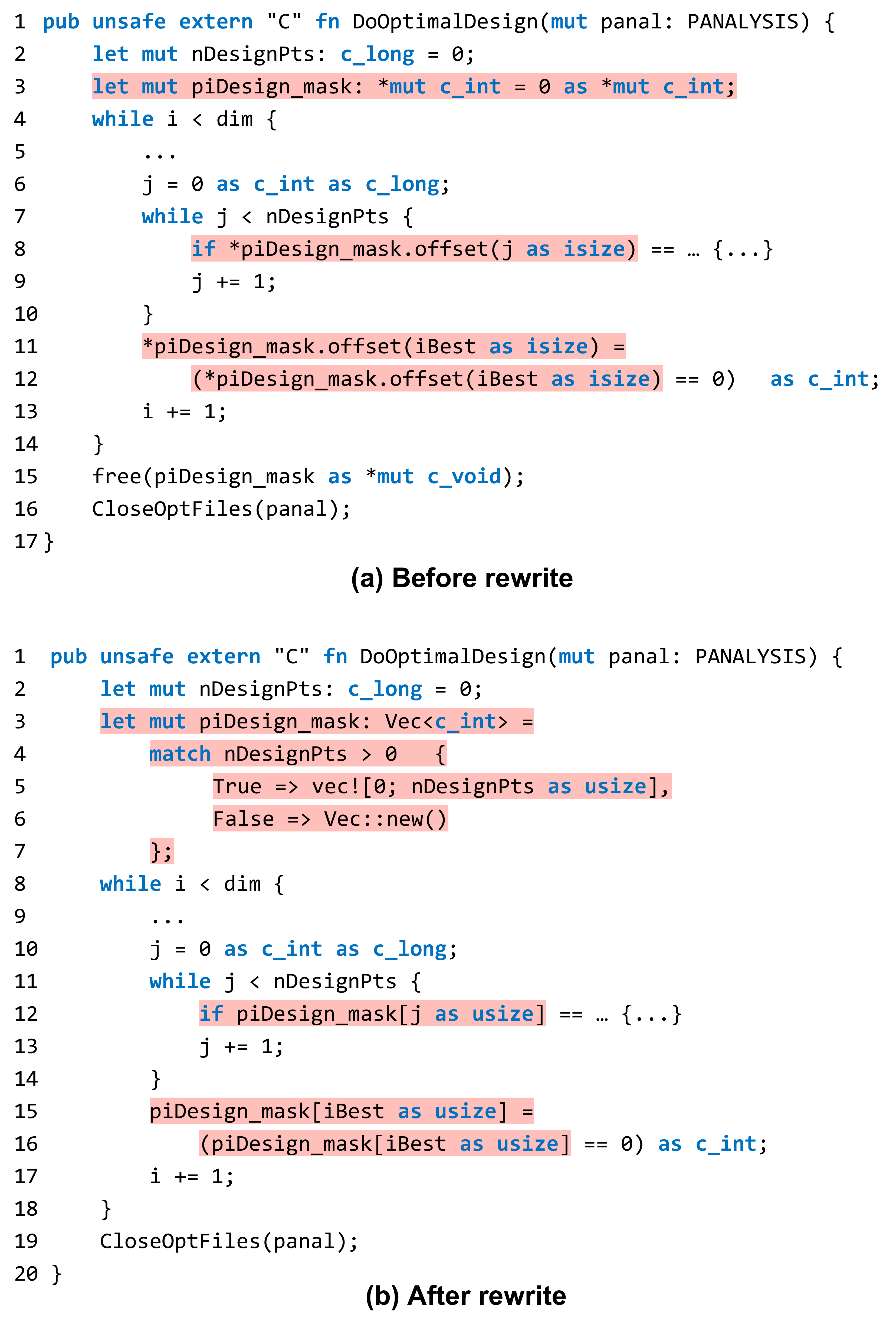}
    \vspace{-6mm}
	\caption{An example of a raw pointer eliminated by the full-featured \toolname\ but not by the ablation NT}
	\label{fig:caseRQ4}
	\vspace{-3mm}
\end{figure}

Figure~\ref{fig:caseRQ4} presents an example that is successfully handled by the full-featured \toolname\ but not by the ablation \textsc{NT}. 
For illustration, we show only a partial snippet of the function \texttt{DoOptimalDesign}, which in its entirety spans 208 lines—posing substantial challenges for the LLM to reason about raw pointer usage.
Guided by the prompt structure shown in Figure~\ref{fig:analysis}, \toolname\ is able to identify the ownership property, the buffer shape, and the memory read/write operations of the raw pointer \texttt{piDesign\_mask}.
These analyses enable the model to lift the pointer to a \texttt{Vec}, following the decision tree defined in the tree in Figure~\ref{fig:tree}.
In contrast, without decision tree-based prompting, the LLM fails to identify this rewriting opportunity within the allowed rewriting budget.

\newpage

\onecolumn
\subsection{Full Prompts of \toolname}
In this section we present prompts of \toolname.  

\begin{promptbox}[Full prompts for pointer lifting]
\small
    Here is a Rust function translated from C code and a raw pointer in it: 

Function: \#FUNCTION\_CONTEXT\#

Raw pointer: \#POINTER\_DECLARATION\#

Please give a retype option for the pointer based on the following decision tree:

(0) Is this pointer belonging to one of the following cases: 
        a. A pointer with a type of void, which means this pointer may have variable type;
        b. A pointer that points to pointers(double pointer or higher level pointers);
 If this pointer belongs to above cases, return with ``CANNOT\_REWRITE'', otherwise, jump to (1)
 
(1) Is the memory value represented by right hand side(RHS) will be owned by this pointer? Typically, if RHS statements are function calls(could be GNU C library functions or other functions seen previously in this program) allocating memory region or initializing objects or arrays, it is regarded as owning pointers, otherwise the pointers should be considered as non-owning, If owning, jump to (2), non-owning, jump to (3)
    
(2) Is the memory an individual region or an array? individual then jump to (4), array jump to (5).

(3) Is the memory is an individual region or an array? individual then jump to (6), array jump to (7).

(4) This is a final state of the decision tree, The pointer seems to be proper to convert to Option<Box<ORIG\_TY{>}{>} by Box::from\_raw, and if this pointer is returned, use Box::into\_raw to convert back.
    
(5) This is a final state of the decision tree, The pointer seems to be proper to convert to Option<Vec<ORIG\_TY{>}{>}.

(6) Will the memory region be written in this function? Yes, jump to (8), No jump to (9)

(7) Will the memory region be written in this function? Yes, jump to (10), No jump to (11)

(8) This is a final state of the decision tree, The pointer seems to be proper to convert to 

Option<\&mut ORIG\_TY>

(9) This is a final state of the decision tree, The pointer seems to be proper to convert to 

Option<\&ORIG\_TY>

(10) This is a final state of the decision tree, The pointer seems to be proper to convert to 

\&mut [ORIG\_TY]

(11) This is a final state of the decision tree, The pointer seems to be proper to convert to 

\&[ORIG\_TY]

You are expected to ONLY return one of following 6 candidates: `Option<Box<ORIG\_TY{>}{>}', `Option<Vec<ORIG\_TY{>}{>}', `Option<\&mut ORIG\_TY>', `Option<\&ORIG\_TY>', `\&mut [ORIG\_TY]', `\&[ORIG\_TY]',
    DO NOT rewrite the function now before more examples given.
\end{promptbox}

% \onecolumn
\begin{promptbox}[Full prompts for statement refactoring]
\small
Good, now based on previous analysis, please rewrite ``Raw pointer'' in ``Function'' below to safe structure you found just now. If the pointer is null, use None or zero length slice to rewrite.

Function: \#FUNCTION\_CONTEXT\#

Raw pointer: \#POINTER\_DECLARATION\#

Here are some examples that may be helpful: 
\#EXAMPLES\#

Following are some struct definitions and types of static variables that may be useful(DO NOT MODIFY ANYTHING IN THEM):

Structs: \#STRUCTS\_USED\#

Statics: \#STATICS\_USED\#

Format instructions: 

1) Your response should include a rust code snippet starting with \texttt{\`{}\`{}\`{}rust} and ending with \texttt{\`{}\`{}\`{}}, which contains the rewritten function inside.

2) When rewriting, please DO NOT change types of function parameters or return values or struct field definitions.

3) Only generate rewritten function and DO NOT generate any Rust `use xxx;' statements outside rewritten function.

4) NEVER omit any lines in ``Function'' even if they remain unchanged.

5) DO NOT omit any variable type annotations.

\end{promptbox}

\begin{promptbox}[Prompts for syntax error fixing]
\small
Please fix the syntax error called ``\#ERROR\_DESCRIPTION\#'', the error is reported at site

\#ERROR\_SITE\#. Also, do not generate any explanations besides fixed code.  
If you cannot fix it due to constraints, response with CANNOT\_FIX. the key context snippet around the error is: 
Snippet: 
\#FOCUS\_SNIPPET\#
You need to fix the error in ``Snippet'' and return the revised snippet. DO NOT return full function.
\end{promptbox}

\begin{promptbox}[Prompts for testcase error fixing]
\small
Good. Previous rewrite has passed the compiler checks successfully, but it failed to pass test cases provided in the program with following stack traces:

Backtrace: \#EXEC\_LOG\#

Current state of the key function in this error is shown below:

line \#START\_LINE\#: \#REWRITTEN\_CODE\#

Originally this function is:
\#ORIGINAL\_CODE\#

The diff of function:
\#DIFF\_LOG\#

Can you try to fix this given error without introducing new raw pointers? Your response should only be a revised version of ``Current Function''. If you think the semantics cannot be rewritten without introducing raw pointers, reply with CANNOT\_FIX.
\end{promptbox}
% \section{Example Appendix}
% \label{sec:appendix}

% This is an appendix.

\end{document}